\documentclass[11pt,letterpaper]{article}

\usepackage{tikz}

\usepackage{mathrsfs}           
\usepackage{vmargin,fancyhdr}   
\usepackage{algorithm}
\usepackage{algorithmic}

\usepackage{enumerate}

\usepackage{amsmath,amssymb}    
\usepackage{verbatim}           
\usepackage{xspace}             
\usepackage{graphicx,float}     
\usepackage{ifthen,calc}        
\usepackage{textcomp}           
\usepackage{fancybox}           
\usepackage{hhline}             
\usepackage{float}              

\usepackage[rflt]{floatflt}     
\usepackage[small,compact]{titlesec}
\usepackage{setspace}
\usepackage{subfigure}

\usepackage{color}
\definecolor{ForestGreen}{rgb}{0.1333,0.5451,0.1333}
\usepackage[letterpaper,dvips,
            colorlinks,linkcolor=ForestGreen,citecolor=ForestGreen,
            backref,
            bookmarks,bookmarksopen,bookmarksnumbered]
           {hyperref}
\usepackage[dvips]{thumbpdf}

\setpapersize{USletter}
\setmarginsrb{.75in}{.5in}        
             {.75in}{.5in}        
             {.25in}{.25in}     
             {.25in}{.5in}      
\newcommand{\showccc}[0]{0}
\newcommand{\ccc}[2][nothing]{
  \ifthenelse{\showccc=0}{}{
    \ensuremath{^{\Lsh\Rsh}}\marginpar{\raggedright\tiny\textsf{%
        \ifthenelse{\equal{#1}{nothing}}{}{\textbf{#1}\\}#2}}}}

\newcounter{hours}\newcounter{minutes}
\newcommand{\hhmm}{%
  \setcounter{hours}{\time/60}%
  \setcounter{minutes}{\time-\value{hours}*60}%
  \ifthenelse{\value{hours}<10}{0}{}\thehours:%
  \ifthenelse{\value{minutes}<10}{0}{}\theminutes}
\ifthenelse{\showccc=0}{\rhead{}}{\rhead{\today \ [\hhmm]}}
\newtheorem{theorem}{Theorem}[section]

\newtheorem{corollary}[theorem]{Corollary}

\newtheorem{lemma}[theorem]{Lemma}
\newtheorem{fact}[theorem]{Fact}
\newtheorem{observation}[theorem]{Observation}

\newcommand{\Proof}[0]{\smallskip\noindent\textit{\textbf{Proof}}\quad}
\newcommand{\Proofof}[1]{\smallskip\noindent\textit{\textbf{Proof of #1:}}\quad}

\newcommand{\QED}[0]{\hfill\ensuremath{\blacksquare}\medspace}
\newcommand{\QEDpart}[1]{\hfill\ensuremath{\blacksquare}(#1)\medspace}

\floatstyle{ruled}
\newfloat{algo}{tbp}{lop}
\floatname{algo}{Algorithm}

\DeclareMathOperator{\trace}{\textbf{tr}}

\usepackage{float}
\floatstyle{plain}\newfloat{myfig}{t}{figs}[section]
\floatname{myfig}{\textsc{Figure}}
\floatstyle{plain}\newfloat{myalg}{H}{algs}[section]
\floatname{myalg}{}
\newcommand{\numiter}{N}

\newcommand{\energy}{\mathcal{E}}
\newcommand{\burn}{\energy}
\newcommand{\power}{\energy}
\newcommand{\epotential}{\burn}
\newcommand{\eflow}{\power}
\newcommand{\econductance}{\mathcal{C}_{eff}}

\newcommand{\approxscale}{\tilde{\lambda}}
\newcommand{\optscale}{\bar{\lambda}}

\newcommand{\optpotentialunscaledv}{\bar{\textbf{x}}}

\newcommand{\approxpotentialunscaledv}{\tilde{\textbf{x}}}

\newcommand{\potentialvert}{\phi}
\newcommand{\potentialvertv}{\phi}

\newcommand{\optpotentialvertv}{\bar{\phi}}

\newcommand{\approxpotentialvertv}{\tilde{\phi}}

\newcommand{\ohmicflow}{\hat{\textbf{f}}}
\newcommand{\ohmicflowv}{\hat{\textbf{f}}}

\newcommand{\potentialedgev}{\textbf{y}}

\newcommand{\saturation}[3]{\textbf{saturation}_{#1}(#2, #3)} 

\newcommand{\flow}{\textbf{f}}
\newcommand{\flowv}{\textbf{f}}
\newcommand{\optflow}{\bar{\textbf{f}}}
\newcommand{\optflowv}{\bar{\textbf{f}}}
\newcommand{\approxflow}{\tilde{\textbf{f}}}
\newcommand{\approxflowv}{\tilde{\textbf{f}}}

\newcommand{\flowmat}{\textbf{F}}
\newcommand{\demandmat}{\textbf{D}}

\newcommand{\pd}{\textbf{P}}

\newcommand{\poly}{\textbf{poly}}

\newcommand{\approxweight}{\tilde{\textbf{w}}}
\newcommand{\approxweightv}{\tilde{\textbf{w}}}
\newcommand{\weight}{\textbf{w}}

\newcommand{\weightsum}{\bar{\textbf{w}}}

\newcommand{\fail}{\textbf{fail} }

\newcommand{\demand}{\textbf{d}}
\newcommand{\demandv}{\textbf{d}}

\newcommand{\edgevertexglobal}{\Gamma}
\newcommand{\edgevertex}{\textbf{B}}
\newcommand{\laplacianglobal}{\mathcal{L}}
\newcommand{\laplacian}{\textbf{L}}

\newcommand{\indicator}{\chi}
\newcommand{\zerosv}{\textbf{0}}
\newcommand{\onesv}{\textbf{1}}
\newcommand{\identity}{\textbf{I}}
\newcommand{\zerosm}{\textbf{0}}

\newcommand{\vecx}{\textbf{x}}

\newcommand{\vecv}{\textbf{v}}

\newcommand{\capacity}{\textbf{u}}
\newcommand{\capacityv}{\textbf{u}}

\newcommand{\maxratio}{U}

\newcommand{\sign}{\textbf{s}}
\newcommand{\signv}{\textbf{s}}
\newcommand{\signlist}{\mathcal{S}}
\newcommand{\optsign}{\bar{\textbf{s}}}
\newcommand{\optsignv}{\bar{\textbf{s}}}

\newcommand{\veca}{\textbf{a}}
\newcommand{\vecb}{\textbf{b}}

\newcommand{\mata}{\textbf{A}}

\newcommand{\matq}{\textbf{Q}}
\newcommand{\matw}{\textbf{W}}
\newcommand{\matm}{\textbf{M}}
\newcommand{\matsumm}{\bar{\textbf{M}}}

\newcommand{\flowvalue}{\mathcal{F}}
\newcommand{\flowvaluev}{\mathcal{F}}
\newcommand{\optflowvaluev}{\bar{\mathcal{F}}}

\title{
Faster Approximate Multicommodity Flow\\
Using Quadratically Coupled Flows
\thanks{Partially supported by the National Science Foundation under grant number CCF-1018463.}}

\author{
  Jonathan A. Kelner \thanks{Partially supported by NFS Awards 0843915 and 1111109}\\
  MIT\\
  \texttt{kelner@mit.edu}\\
\and
  Gary L.\ Miller\\
  CMU\\
  \texttt{glmiller@cs.cmu.edu}\\
\and
  Richard Peng \thanks{Supported by a Microsoft Fellowship}\\
  CMU  \thanks{Part of this work was done while at Microsoft Research New England}\\
  \texttt{yangp@cs.cmu.edu}\\
}

\begin{document}

\maketitle

\begin{abstract}
The maximum multicommodity flow problem is a natural generalization of the maximum
flow problem to route multiple distinct flows.
Obtaining a $1-\epsilon$ approximation to the multicommodity flow problem on graphs
is a well-studied problem.
In this paper we present an adaptation of recent advances in single-commodity
flow algorithms to this problem.
As the underlying linear systems in the electrical problems of multicommodity flow
problems are no longer Laplacians, our approach is tailored to generate specialized
systems which can be preconditioned and solved efficiently using Laplacians.
Given an undirected graph with $m$ edges and $k$ commodities, we give algorithms
that find $1-\epsilon$ approximate solutions to the maximum concurrent flow problem
and the maximum weighted multicommodity flow problem
in time $\tilde{O}(m^{4/3}\poly(k,\epsilon^{-1}))$
\footnote{We use $\tilde{O}(f(m))$ to denote
$\tilde{O}(f(m) \log^c f(m))$ for some constant $c$}.
\end{abstract}

\thispagestyle{empty}
\newpage
\setcounter{page}{1}
\section{Introduction}
\label{sec:intro}

The multicommodity flow problem is a natural extension of the
maximum flow problem.
One of its variations, maximum concurrent multicommodity flow,
asks to route multiple demands simultaneously in a network
subject to capacity constraints.
In this setting we're given an undirected, capacitated graph
$G = (V, E, \capacityv)$ where $\capacityv : E \rightarrow \Re^+$,
and $k$ source-sink pairs $(s_1, t_1) \ldots (s_k, t_k)$.
The goal is to find the maximum $\lambda$ such that
there exist $k$ flows $\flowv_1 \ldots \flowv_k$
where $\flowv_i$ routes $\lambda$ units of flow between $s_i$ and $t_i$,
and the total flow along each edge obey the following capacity constraint:

\begin{align*}
\sum_{i} |\flow_i(e)| \leq \capacity(e) \qquad \forall e \in E
\end{align*}

\subsection{Related Work}
\label{sec:related}

The simplest version of the problem is with two commodities in an
undirected graph.
In this case the problem was shown to be reducible to two single
commodity maximum flow problems \cite{RothschildW1966}.
When there are $3$ or more commodities though, this connection no longer
holds and all of the (almost) exact algorithms for multicommodity flow problems
involve solving a linear programming formulation.
For these linear programs, the method with the best asymptotic behavior
is the interior point algorithm, which requires solving $O(m^{1/2})$ linear systems.
By tracking inverses of these systems and making low rank updates, Vaidya
showed an algorithm with running time $\tilde{O}(k^{5/2}m^{3/2}n)$ \cite{Vaidya89}.
This is not very far from the natural barrier of $\tilde{\Omega}(k^{1/2}m^{1/2}n^2)$
for this type of approach, which arises from the need to compute a dense matrix-vector
product involving the inverse in each iteration.

Subsequent work on multicommodity flow focused on obtaining $1+\epsilon$
approximate solutions in faster time.
This work initially focused on the case of small $k$, and the algorithms
are based on solving multiple minimum cost flow problems \cite{LeightonMPSTT91}.
When combined with the minimum cost flow algorithm from \cite{DaitchS08},
these algorithms gave a running time of $\tilde{O}(m^{1.5}\poly(k, \epsilon^{-1}))$.

More recent approaches have favored using a less expensive inner loop to obtain
better bounds when $k$ is large.
Specifically multiplicative weights update method  using single source shortest
path routines as oracles \cite{GargKonemann, Fleischer00}.
The most recent among these approaches obtained a running time of
$\tilde{O}(nm/\epsilon^2)$ using dynamic graph data structures\cite{Madry10}.
These methods give better performance the case where $k$ is large.
However, when applied to instances with a smaller value of $k$ these approaches
encounter similar issues to those encountered by path based single-commodity
flow algorithms: the flow decomposition barrier at $\Omega(nm)$.
This barrier stems from the fact that if we decompose a flow into a list of paths
of paths, the total size of these paths can be $\Omega(nm)$.

An alternate approach to solving linear systems, has led to a possible
way to circumvent both the dense inverse and flow decomposition barrier.
The graph-like nature of the underlying linear system, graph Laplacians, allows one to
find sparse approximations of it, which, when used to precondition iterative solvers,
led to speedups in solving such systems \cite{Vaidya91}.
This approach using graph Laplacians has been extended greatly in subsequent works,
leading to algorithms with nearly-linear running times \cite{SpielmanTengSolver}.
Graph Laplacians are closely connected with problems involving a single flow,
such as maximum flow, minimum cost flows and shortest path.
To the best of our knowledge this connection was first observed in
\cite{DaitchS08}, leading to, among others, a faster algorithm for
minimum cost flow.
This algorithm is used as a subroutine in the
$\tilde{O}(m^{1.5}\poly(k, \epsilon^{-1}))$ algorithm mentioned earlier.

\subsection{Our Work}
\label{sec:our}

Recently, the running time for approximate maximum flow in undirected
graphs has been improved to $\tilde{O}(m^{4/3})$ \cite{ChristianoKMST10}.
A natural question arising from this is whether this algorithm can also be
extended to multicommodity flow.
Even though in the $2$-commodity setting their algorithm can be invoked in a
black-box manner \cite{RothschildW1966}, a more general examination of this
setting is helpful for understanding the main components of our extensions.

In order to further simplify the $2$-commodity case,
we assume that each edge have unit capacity.
That is, we want to find two flows $\flowv_1$ and $\flowv_2$ that
meet their respective demands, and satisfy the following capacity
constraint on each edge $e$.

\begin{align*}
|\flow_1(e)| + |\flow_2(e)| \leq 1
\end{align*}

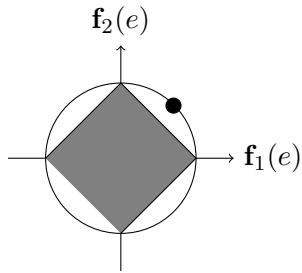
\begin{figure}[h]
\begin{center}
\begin{tikzpicture}
    \draw[->] (-1.5,0) -- (1.5,0) node[right] {$\flowv_1(e)$};
    \draw[->] (0,-1.5) -- (0,1.5) node[above] {$\flowv_2(e)$};
    \draw { (0,-1) -- (1,0) -- (0, 1) -- (-1, 0) } [fill = gray];
    \draw (0, 0) circle (1);
    \draw (0.70, 0.70) circle (0.1) [fill = black];
\end{tikzpicture}
\end{center}
\caption{Unit $l_1$ ball representing region of feasible flows (gray),
with unit $l_2$ ball being the quadratic capacity constraint.
The point $(\flow_1(e) = \sqrt{2}/2, \flow_2(e) = \sqrt{2}/2)$
obeys this constraint, but exceeds the edge's capacity}
\label{fig:unitball}
\end{figure}

One way to visualize this constraint is by considering each flow assignment
as a coordinate in the $2$-D plane.
Then a point $(\flowv_1(e), \flowv_2(e))$ obeys this constraint if it's inside
the unit $L_1$ ball, as shown in Figure \ref{fig:unitball}.
The Christiano et al. algorithm \cite{ChristianoKMST10} produces a flow
that approximately satisfies edge capacities by solving a series of electrical problems.
To form such electrical problems, they assign one
resistor per edge, leading to a quadratic term of the form $\flow(e)^2$.
Since $\flow(e) \leq 1$ is equivalent to $\flow(e)^2 \leq 1$, they're able
to bound the energy of each of these terms by $1$.
A natural generalization to two commodities would be to bound the
sum of the squares of the two flows,
leading to an instance of what we define as \textbf{Quadratically Capacitated Flows}.
Specifically we would like the following to hold along each edge:

\begin{align*}
\flow_1(e)^2 + \flow_2(e)^2 \leq 1
\end{align*}

Note that the flow $(\flow_1(e), \flow_2(e)) = (1, 0)$ has energy $1$,
so the RHS value of $1$ is tight.
This corresponds to allowing any $(\flowv_1(e), \flowv_2(e))$ that's within the unit
$L_2$ ball, and as shown in Figure \ref{fig:unitball}.
However, in the $2$-commodity case it's possible for flow settings that over-congest
the edge to still satisfy this energy constraint.
In other words,  it's possible for a point to be in the unit $L_2$ ball
but outside the unit $L_1$ ball.
For example, the flow $(\flow_1(e) = \sqrt{2}/2, , \flow_2(e) = \sqrt{2}/2)$
also meets this energy constraint despite having a congestion $\sqrt{2}$.

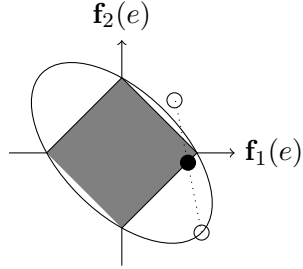
\begin{figure}[h]
\begin{center}
\begin{tikzpicture}
    \draw[->] (-1.5,0) -- (1.5,0) node[right] {$\flowv_1(e)$};
    \draw[->] (0,-1.5) -- (0,1.5) node[above] {$\flowv_2(e)$};
    \draw { (0,-1) -- (1,0) -- (0, 1) -- (-1, 0) } [fill = gray];
    \draw[rotate = -45] (0, 0) ellipse (1.5 and 0.8);
    \draw (0.70, 0.70) circle (0.1);
    \draw (1.06, -1.06) circle (0.1);
    \draw[dotted] (1.06, -1.06) -- (0.71, 0.71);
    \draw (0.88, -0.13) circle (0.1) [fill = black];
\end{tikzpicture}
\end{center}
\caption{Modified energy constraint after a flow of
$(\sqrt{2}/2, \sqrt{2}/2)$ has been added.
Note that the point returned is still outside of the
unit $l_1$ ball, but the average is within it.}
\label{fig:quadraticallycoupled}
\end{figure}

One possible remedy to this problem is to introduce more intricate
quadratic coupling between the two commodities.
However, we still need to set the constraint so that any flow whose total
congestion is below the capacity falls within this ellipse.
This is equivalent to the ellipse containing the unit $L_1$ ball, which along
with the fact that ellipses have smooth boundaries means we must allow
some extra points.
For example, the modified ellipse in Figure \ref{fig:quadraticallycoupled} once
again allows the returned solution to have large congestion.
In general, computing a single quadratically capacitated flow
can lead to the edge being over-congested by a factor of $\sqrt{k}$,
giving a $\sqrt{k}$ approximation.

As a result, instead of computing a single quadratically capacitated flow,
we compute a sequence of them and \textbf{average} the result.
Note that although both $(\flow_1(e) = \sqrt{2}/2, , \flow_2(e) = \sqrt{2}/2)$
and $(\flow_1(e) = \sqrt{2}/2, , \flow_2(e) = -\sqrt{2}/2)$
have congestion of $\sqrt{2}$, if we average them we're left with a flow
with congestion only $\sqrt{2}/2$.
If we're only concerned with keeping the average congestion small,
the flows computed in previous iterations can give us some 'slack' in
certain directions.
As it turns out, if we compute the coupling matrix based on the flows returned
so far, it's possible to move the average gradually get closer to the unit $L_1$ ball.
For example, the average of the two flows returned in Figure \ref{fig:quadraticallycoupled}
is inside the unit $L_1$ ball despite both falling outside of it.

The problem now becomes finding feasible quadratically capacitated flows.
The Christiano et al. algorithm \cite{ChristianoKMST10} can be adapted
naturally to this problem,
providing that we can find flows that minimizes a weighted sum of the energy terms.
We define this generalization of electrical flows as \textbf{quadratically coupled flows}.
Just like their single commodity version, the minimum energy quadratically coupled
flows can also be computed by solving linear systems.

The remaining difficulty of the problem is now with solving these linear systems.
Most of the combinatorial preconditioning framework relies on the system
being decomposable into $2$-by-$2$ blocks corresponding to single edges.
For quadratically coupled flows, the resulting systems are only decomposable
into $2k$-by-$2k$ blocks.
These systems are also encountered in stiffness matrices of finite element
systems \cite{BHV04}, and $k$-dimensional trusses \cite{DaitchS07, AT11}.
To date a nearly-linear time solver that can handle all
such systems remains elusive.

Instead of solving these systems directly, we show,
by more careful analysis of our algorithm
that generates the quadratically capacitated flow problems,
that it suffices to consider a more friendly subset of them.
It can be shown that the $\sqrt{k}$ factor deviation that occurs in the
uncoupled electrical problem is also the extent of our loss
if we try to approximate these more friendly
quadratically coupled flows with uncoupled ones.
However, in the quadratic case such losses are fixable using
preconditioned iterative methods, allowing us to solve the systems
arising from quadratically coupled flows by solving a number of
graph Laplacians instead.
This leads us our main result, which can be stated as:

\begin{theorem}
Given an undirected, capacitated graph $G = (V, E, \capacityv)$
with $m$ edges, along with $k$ commodities and their demands
$\demandv_1 \ldots \demandv_i$.
There is an algorithm that computes an $1-\epsilon$ approximate
maximum  concurrent flow in time:

\begin{align*}
\tilde{O}(m^{4/3}\poly(k, \epsilon^{-1}))
\end{align*}

\end{theorem}

An overview of the main steps of the algorithm is shown in
Section~\ref{sec:overview}.
Our approach also extends to maximum weighted multicommodity flow,
which we show in Appendix \ref{sec:weighted}.

\section{Preliminaries}
\label{sec:formulations}


The maximum concurrent multicommodity problem concerns the simultaneous routing
of various commodities in a capacitated network.
For our purposes, the graph is an undirected, capacitated graph
$G=(V, E, \capacityv)$ where $\capacityv : E \rightarrow R^{+}$
is the capacity of each edge.
If we assign an arbitrary orientation to the edges, we can denote
the edge-vertex incidence matrix $\edgevertex \in \Re^{m \times n}$ as:

\begin{align}
\label{definition:edgevertex}
\edgevertex(e, u) =
\left\{
\begin{array}{lr}
1 & \text{if u is the head of e} \\
-1 & \text{if u is the tail of e} \\
0 & \text{otherwise}
\end{array}
\right.
\end{align}
Then, for a (single commodity) flow $\flowv$, the excess of the flow at each vertex is given by the
length n vector $\edgevertex^T \flowv$.

Throughout the paper we let $k$ be the number of commodities routed.
It can be shown that it suffices to solve the $k$-commodity flow problem
for fixed vertex demands
$\demandv_1, \demandv_2 \ldots \demandv_k$ one for each commodity.
The goal of finding a flow that concurrently routes these demands in turn becomes
finding $\flowv_i$ for each commodity such that:

\begin{align}
\edgevertex^T \flowv_i = \demandv_i
\end{align}

The other requirement for a valid flow is that the flows cannot exceed the
capacity of an edge.
Specifically we need the following constraint for each edge $e$:

\begin{align}
\sum_{1 \leq i \leq k} |\flow_i(e)| \leq \capacity(e)
\end{align}

\subsection{Notations for $k$-Commodity Flow and Vertex Potentials}

The extension of a single variable indicating flow/vertex potential on
an edge/vertex to $k$ variables creates several notational issues.
We use a length $km$ vector $\flowv \in \Re^{km}$ to denote a k-commodity flow,
and allow for two ways to index into it based on commodity/edge respectively.
Specifically, for a commodity $i$, we use $\flowv_i$ to denote the length $m$
vector with the flows of commodity $i$ along all edges and
for an edge $e$, we use $\flowv(e)$ to denote the length $k$ vector containing
the flows of all $k$ commodities along this edge.

This definition extends naturally to vectors over all (vertex, commodity) pairs as well.
We let $\demandv \in \Re^{nk}$ be the column vector obtained by concatenating the length  $k$ demand vector over all
$n$ vertices.  
If the edges are labeled $e_1 \ldots e_m$ and the vertices $v_1 \ldots v_n$, then $\flowv$
and $\demandv$ can be written as:

\begin{align}
\flowv^T = & \left[ \flowv(e_1)^T, \flowv(e_2)^T, \ldots , \flowv(e_m)^T \right] \\
\demandv^T = & \left[\demand(v_1)^T, \flowv(v_2)^T, \ldots , \demandv(v_n)^T \right]
\end{align}

We can also define larger matrices that allows us to express these conditions across
all $k$ commodities, and their interactions more clearly.
The edge-vertex incidence matrix that maps between $\demandv$ and $\flowv$ is
the Kronecker product between $\edgevertex$ and the $k \times k$ identity matrix $\identity_k$.

\begin{align}
    \edgevertexglobal = & \edgevertex \otimes \identity_k
\end{align}

and $\flowv$ meeting the demands can be written as:

\begin{align}
    \edgevertexglobal^T \flowv = \demandv
\end{align}

Note that flows of two commodities passing in opposite directions through the
edge do not cancel each other out.

By obtaining crude bounds on the flow value using bottle neck shortest paths
and binary searching in the same way as in \cite{ChristianoKMST10},
the maximum concurrent multicommodity flow
problem can be reduced to $O(\log{n})$ iterations of checking whether
there is a $k$-commodity flow $\flowv$ that satisfies the following:

\begin{align*}
||\flow(e)||_1
= \sum_{i = 1}^k |\flow_i(e)| \leq & u(e) \qquad \forall e \in E\\
\edgevertexglobal \flowv = & \demandv
\end{align*}

\subsection{Quadratic Generalizations}

We  define two generalizations of electrical flows to multiple commodities.
Our main goal is to capture situations where the amount of flows of one type allowed on an
edge depends inversely on the amount of another flow, so the flows are  ``coupled.''
To do so, we introduce a positive-definite, block-diagonal matrix $\pd \in \Re^{km \times km}$
such that $\pd = \sum_e \pd(e)$ and each $\pd(e)$ is a $k \times k$ positive definite
matrix defined over the $k$ entries corresponding to the flow values on edge $e$.

For each edge the $k$ flows along an edge $e$, $\flowv(e)$ we get a
natural quadratic penalty or energy dissipation term:

\begin{align}
\power_{\flowv}(\pd, e) = \flow(e)^T \pd(e) \flow(e)
\end{align}

Summing these gives the total energy dissipation of a set of flows, denoted using $\power$.

\begin{align}
\power_{\flowv}(\pd)
= & \sum_e \power_{\flowv}(\pd, e) \nonumber \\
= & \flowv^T \pd \flowv
\end{align}

In the  {\bf Quadratically Coupled Flow} problem, we aim to find  a flow $\flowv$ that satisfies
all of the demand constraints and minimizes the total energy dissipation, namely:

\begin{align}
\min \qquad & \eflow_{\flowv}(\pd)\\
\text{subject to:} \qquad & \edgevertexglobal^T \flowv = \demandv
\end{align}

The minimum is denoted by $\eflow(\pd)$.
We can define the related potential assignment problem, where the goal is to assign
potentials to the vertices to separate the demands.
Note that due to there being $k$ commodities, a potential can be assigned to each
 (flow, vertex) pair, creating $\potentialvertv \in \Re^{kn}$.
This vector can also be viewed as being composed of $n$ length $k$ vectors,
with the vector at vertex $u$ being $\potentialvertv(u)$.
Given an edge $e = (u,v)$ whose end points connects vertices with potentials
$\potentialvertv(u)$ and $\potentialvertv(v)$, the difference between its
end points is $\potentialvertv(u) - \potentialvertv(v)$.
In order to map this length $nk$ vector into the same support as the k-commodity
flows along edges, we need to multiply it by $\edgevertexglobal$,
and we denote the resulting vector as $\potentialedgev$:

\begin{align}
\potentialedgev = & \edgevertexglobal \potentialvertv
\end{align}
The energy dissipation of an edge with respect to $\potentialvertv$ can in turn be defined as:

\begin{align}
\burn_\potentialvertv(\pd, e)
= & \potentialedgev \pd(e)^{-1} \potentialedgev
\end{align}

Note that this definition relies on $\pd(e)$ being positive definite and therefore invertible
on the support corresponding to edge $e$.
This can in turn be extended analogously to the energy dissipation of a set of potentials as:

\begin{align}
\epotential_\potentialvertv(\pd)
= & \sum_{e} \burn_\potentialvertv(\pd, e) \nonumber \\
= & \potentialvertv^T \edgevertexglobal^T \pd^{-1} \edgevertexglobal \potentialvertv
\end{align}

We further generalize the definition of a Laplacian to $k$ commodities:

\begin{align}
  \laplacianglobal = \edgevertexglobal^T \pd^{-1} \edgevertexglobal
\end{align}
Thus, the energy dissipation of a set of potentials also equals to $\potentialvert^T \laplacianglobal \potentialvert$.
Which leads to the following maximization problem, which is the dual of the quadratically coupled
flow problem.

\begin{align}
\max \qquad & (\demandv^T \potentialvertv)^2\\
\text{subject to:} \qquad & \epotential_\potentialvertv(\pd) \leq 1
\end{align}

We denote the optimum of this value using $\econductance(\pd)$ and will show in
Section \ref{sec:electricalproperties} that $\econductance(\pd) = \eflow(\pd)$.

Another coupled flow problem that's closer to the maximum concurrent flow problem
is one where we also bound the {\bf saturation} of them w.r.t. $\pd$,
where {\bf saturation} is the square root of the energy dissipation.

\begin{align}
\saturation{\flow}{\pd}{e} =\sqrt{\flowv(e)^T \pd(e) \flowv(e)}
\end{align}

Finding a flow with bounded saturation per edge will be then called the
{\bf Quadratically Capacitated Flow} problem.

\section{Overview of Our Approach}
\label{sec:overview}

A commonality of the algorithms for flow with Laplacian solves as an inner loop
\cite{DaitchS08, ChristianoKMST10} is that they make repeated computations of
an optimum electrical flow in a graph with adjusted edge weights.
The main problem with extending these methods to $k$-commodity
flow is that the Laplacian for $k$-commodity electrical flow $\laplacianglobal$
is no longer symmetrically diagonally dominant.
Our key observation in resolving this issue is that when the energy matrices
$\pd(e)$ are well-conditioned, we can precondition $\pd$ with a diagonal matrix.
Then using techniques similar to those in \cite{BHV04},
we can solve systems involving $\laplacianglobal$ using
a small number of Laplacian linear system solves.
Our algorithm for $k$-commodity flow has the following layers with a description
in Figure~\ref{fig:structurefigure} as well.

\begin{enumerate}
\item We adapt the algorithm from \cite{LeightonMPSTT91} to use flows with electrical
capacity constraints associated with the $k$ commodities instead of minimum cost
flow as its oracle call.
  At the outermost level, the approximately multi-commodity flow
  algorithm, repeatedly computes a positive definite matrix $\pd(e)$
  for each edge based on the flows on it so far on that edge, and boost their
  diagonal entries to keep their condition number at most $\poly(k)$.
  The outermost level then calls an algorithm that computes quadratically capacitated
  flow that is:

\begin{align}
\saturation{\approxflowv}{\pd}{e}
\leq & \max_{\flowv, ||\flowv(e)||_1 \leq \capacity(e)} \saturation{\flowv}{\pd}{e}
\end{align}

After repeating this process $\poly(k)$ times, averaging these flows gives one where
$||\approxflowv(e)||_1 \leq (1 + \epsilon) \capacity(e)$ on all edges.
We give two methods for computing $\pd$ in Sections \ref{sec:outer} and \ref{sec:outer1}.

\item We use an algorithm that's a direct extension of the electrical flow based
maximum flow algorithm from \cite{ChristianoKMST10} to minimize the maximum saturation
of an edge.
This stage of the algorithm in turn solves $\tilde{O}(m^{1/3})$ quadratically
 coupled flows where the energy coupling on an edge is $\tilde{\pd}(e) = \weight_e \pd(e)$.
Note that since $\pd(e)$ was chosen to be well conditioned and $\weight_e$ is a scalar,
the $\tilde{\pd}(e)$s that we pass onto the next layer on remains well-conditioned.
This is presented in Section~\ref{sec:inner}.

\item In turn the Quadratically Capacitated Flow Algorithm makes calls to an
algorithm that computes a quadratically coupled flow.
The almost-optimal quadratically coupled flow is obtained by linear solves involving $\laplacianglobal$.
Specifically, we show that preconditioning each $P(e)$ with a diagonal matrix allows
us to decouple the $k$-flows, at the cost of a mild condition number set in the outermost layer.
Then using preconditioned Chebyshev iteration, we obtain an almost optimal
 quadratically coupled flow using $\poly(k, \epsilon^{-1})$ Laplacian solves on a matrix
with $m$ non-zero entries.
Properties of the $k$-commodity electrical flow, as well as bounds on the error and
convergence of the solves are shown in Section \ref{sec:electrical}.

\end{enumerate}

\begin{figure}

\begin{itemize}
\item[] \textsc{MaxConcurrentFlow}\\
{\bf Constraint on desired flow:} For each $e$, total flow $\vert f_e \vert\leq 1$.\\
Repeatedly updates energy matrices using matrix multiplicative weights.  Makes $\mathrm{poly}(k)$ oracle calls to:
\begin{itemize}
\item[] \textsc{QuadraticallyCapacitatedFlow}\\
{\bf Constraint on desired flow:}  For each $e$, $\saturation{\flowv}{\pd}{e}=\sqrt{\flowv(e)^T \pd(e) \flowv(e)} \leq 1$.\\
Repeatedly updates energy matrices using (scalar) multiplicative weights.  Makes $\tilde{O}(m^{1/3})$ oracle calls to:
\begin{itemize}
\item[] \textsc{QuadraticallyCoupledFlow}\\
{\bf Constraint on desired flow:}  Minimize total dissipated energy $\eflow_{\flowv}(\pd)=\sum_e \flowv(e)^T \pd(e) \flowv(e).$
Solves 1 linear system using:
\begin{itemize}
\item[] \textsc{PreconCheby} \\
Solves non-Laplacian system by preconditioning with $k$ $n\times n$ Laplacians.  Solves these using $k$ calls to nearly-linear time Laplacian solvers.
\end{itemize}
\end{itemize}
\end{itemize}
\end{itemize}

\caption{The high-level structure of the algorithm and the approximate number of calls made to each routine (for fixed $\epsilon$).}
\label{fig:structurefigure}

\end{figure}

\section{Approximate Computation of Quadratically Coupled Flows}
\label{sec:electrical}

\subsection{Quadratically Coupled Flows and Vertex Potentials}
\label{sec:electricalproperties}

Let $\optpotentialunscaledv$ be the vector such that
$\laplacianglobal \optpotentialunscaledv = \demandv$.
It can be shown that $\epotential(\pd, \potentialvertv)$ is maximized
when $\potentialvertv$ is a multiple of $\optpotentialunscaledv$.
Then the scaling quantity $\optscale$ as well as the optimum set of
potential $\optpotentialvertv$ are:

\begin{align}
\optscale
    = & \sqrt{\demandv^T \laplacianglobal^{+} \demandv} \label{eqn:optscale} \\
\optpotentialvertv
    = & \frac{1}{\optscale} \optpotentialunscaledv \nonumber \\
    = & \frac{1}{\optscale} \laplacianglobal^{+} \demandv \label{eqn:optpotential}
\end{align}

Note that $\demandv$ satisfies $\onesv_i^T \demandv = 0$ for all $1 \leq i \leq k$.
Also, since $\pd$ is positive-semidefinite, the null space of $\laplacianglobal$ is precisely
the space spanned by the $k$ vectors $\onesv_i$.
Therefore $\demandv$ lies completely within the column space of $\laplacianglobal$ and
we have $\laplacianglobal \laplacianglobal^{+} \demandv  = \demandv$.
The value of $\demandv^T \optpotentialvertv$ is then:

\begin{align}
\demandv^T \optpotentialvertv
= & \demandv^T \frac{1}{\optscale} \laplacianglobal^{+} \demandv \nonumber \\
= & \optscale
\end{align}

The optimal quadratically coupled flow can be obtained from the optimal vertex potentials as follows:

\begin{align}
\optflowv
= & \pd^{-1} \edgevertexglobal \optpotentialunscaledv \nonumber \\
= & \optscale \pd^{-1} \edgevertexglobal \optpotentialvertv
\label{eqn:optflow}
\end{align}

We can prove the following generalizations of standard facts about electrical flow/effective resistance
for multicommodity electrical flows.

\begin{fact}
\label{fact:electricalpotential}
\begin{enumerate}
    \item $\optflow$ satisfies the demands, that is $\edgevertexglobal^T \optflowv = \demandv$. \label{part:electricalconservation}
    \item $\eflow(\pd, \optflow) = \demandv^T \laplacianglobal^{+} \demandv$ \label{part:electricalenergy}
    \item For any other flow $\flowv$ that satisfies the demands, $\eflow(\pd, \flowv) \geq \eflow(\pd, \optflowv)$. \label{part:optelectrical}
\end{enumerate}
\end{fact}

\Proof

Part \ref{part:electricalconservation}

\begin{align}
\edgevertexglobal^T \optflowv
= & \optscale \edgevertexglobal^T \pd \edgevertexglobal \optpotentialvertv \nonumber \\
= & \optscale \laplacianglobal
    \frac{1}{\optscale} \laplacianglobal^{+} \demandv \nonumber \\
= & \demandv
\end{align}

Part \ref{part:electricalenergy}

\begin{align}
\eflow(\pd, \optflowv)
= & \optflowv^T \pd \optflowv \nonumber \\
= & \demandv^T \laplacianglobal^{+} \demandv
    (\pd^{-1} \edgevertexglobal \optpotentialvertv)^T \pd (\pd^{-1} \edgevertexglobal \optpotentialvertv) \nonumber \\
= & \demandv^T \laplacianglobal^{+} \demandv
    \optpotentialvertv^T \edgevertexglobal^T \pd^{-1} \pd \pd^{-1} \edgevertexglobal \optpotentialvertv \nonumber \\
= & \demandv^T \laplacianglobal^{+} \demandv \optpotentialvertv^T \laplacianglobal \optpotentialvertv \nonumber \\
= & \demandv^T \laplacianglobal^{+} \demandv
    \quad \mbox{Since $\optpotentialvertv^T \laplacianglobal \optpotentialvertv = 1$}
\end{align}

Part \ref{part:optelectrical}

Let $\flowv$ be any flow satisfying $\edgevertexglobal^T \flowv = \demandv$.
Then we have:
\begin{align}
\eflow(\pd, \flowv)
\geq & (\flowv^T \pd \flowv) (\optpotentialvertv^T \edgevertexglobal^T \pd^{-1} \edgevertexglobal \optpotentialvertv)
    \qquad \mbox{Since $\optpotentialvertv^T \laplacianglobal \optpotentialvertv = 1$} \nonumber \\
= & ||\pd^{1/2} \flowv||_2^2 ||\pd^{-1/2} \edgevertexglobal \optpotentialvertv ||_2^2 \nonumber \\
\geq & (\flowv^T \edgevertexglobal \optpotentialvertv)^2
    \qquad \mbox{By Cauchy-Schwarz inequality} \nonumber \\
= & (\demandv^T \optpotentialvertv)^2
    \qquad \mbox{Since $\flowv$ satisfies the demands} \nonumber \\
= & \eflow(\pd, \optflowv)
    \qquad \mbox{By Part \ref{part:electricalenergy}}
\end{align}

\QED

\subsection{Finding Almost Optimal Vertex Potentials}

The main part of computing an almost optimal set of vertex potentials from
\ref{eqn:optpotential} is the computation of $\laplacianglobal^{+} \demandv$.
Since $\pd$ is no longer a diagonal, the matrix $\laplacianglobal$ is no longer a
Laplacian matrix.
However, in certain more restrictive cases that still suffice for our purposes
we can use lemma $2.1$ of \cite{BHV04}:

\begin{lemma}
\label{lem:precongeneral}
For any matrices $V, G, H$, if $H \preceq G \preceq \kappa H$,
then $VHV^T \preceq G \preceq \kappa VHV^T$.
\end{lemma}

\Proof
Consider any vector $x$, we have:
\begin{align}
x^TVHV^Tx
= &(V^Tx)^T H (V^Tx) \nonumber \\
\preceq &  (V^Tx)^T G (V^Tx)
 = x^T VGV^T x
\end{align}

and

\begin{align}
x^TVGV^Tx
= & (V^Tx)^T G (V^Tx) \nonumber \\
\preceq & \kappa (V^Tx)^T H (V^Tx)
 = \kappa x^T VHV^T x
\end{align}
\QED

This lemma allows us to precondition $\pd$ when each of $\pd(e)$ is well-conditioned,
specifically:

\begin{lemma}
\label{lem:precon}
If there exist a constant $\kappa$ such that for all $e$,
$\kappa \lambda_{\min}(\pd(e)) \geq \lambda_{\max} (\pd(e))$, then we can find
a Laplacian matrix $\tilde{\laplacianglobal}$ such that
$\tilde{\laplacianglobal} \preceq \laplacianglobal \preceq \kappa \tilde{\laplacianglobal}$.
\end{lemma}

\Proof
Consider replacing $\pd(e)$ with $\tilde{\pd}(e) = \lambda_{min}(\pd(e)) \identity(e)$
where $\identity(e)$ is the $k$ - by - $k$ identity matrix.

Then since $\pd(e) \preceq \lambda_{max} \identity(e)$ as well, we have:

\begin{align}
\tilde{\pd}(e)
\preceq \pd(e)
\preceq \lambda_{max}(\pd(e)) \identity(e)
= \frac{\lambda_{max}(\pd(e))}{\lambda_{\min}(\pd(e))} \tilde{\pd}(e)
\end{align}

Then applying Lemma \ref{lem:precongeneral} with $V = \edgevertexglobal^T$,
$G = \pd$ and $H = \tilde{\pd}$ gives:

\begin{align}
\tilde{\laplacianglobal} \preceq \laplacianglobal \preceq \kappa \tilde{\laplacianglobal}
\end{align}

\QED

The following fact then allows us to solve linear equations on $\laplacianglobal$
by solving linear systems on $\tilde{\laplacianglobal}$ instead:

\begin{lemma}
\label{lem:preconcheby}

(preconditioned Chebyshev) \cite{Saad:1996:book, Axe94}
Given matrix $A$, vector $b$, linear operator $B$ and a constant $\kappa$ such that
$B \preceq A^{+} \preceq \kappa B$ and a desired error tolerance $\theta$.
We can compute a vector $x$ such that
$||x - A^{+}b||_{A} \leq \theta ||A^{+}b||_A$
using $O(\sqrt{\kappa}\log{1/\theta})$ evaluations of the linear operators $A$
and $B$.
\end{lemma}

Note that due to $\edgevertexglobal$ being $k$ copies of the edge-vertex incidence matrix,
the matrix $\tilde{L}$ is actually $k$ Laplacians arranged in block-diagonal form.
This allows us to apply SDD linear system solves to apply
an operator that is close to the pseudo-inverse of $\tilde{\laplacianglobal}$,
which we in turn use to solve systems involving $\laplacianglobal$ using Lemma \ref{lem:preconcheby}.

\begin{lemma}
\label{lem:laplaciansolver}
\cite{SpielmanTengSolver, KoutisMP10, KoutisMP11}
Given a Laplacian matrix of the form $\laplacian = \edgevertex^T W \edgevertex$
for some diagonal matrix $W \geq 0$, there is a linear operator $A$ such that

\[
A \preceq L^{+} \preceq 2A
\]

And for any vector $\vecx$, $A \vecx$ can be evaluated in time $\tilde{O}(m)$ where $m$ is
the number of non-zero entries in $\laplacian$.
\end{lemma}

We can now prove the main result about solving systems involving $\laplacianglobal$.

\begin{lemma}
\label{lemma:solve}
Given any set of energy matrices on edges $\pd$ such that
$\lambda_{max}(\pd(e)) \leq \kappa \lambda_{min}(\pd(e))$,
a vector $\demandv$ and error parameter $\delta$.
We can find an almost optimal set of vertex potentials
$\approxpotentialunscaledv$ such that:
\begin{align}
||\approxpotentialunscaledv - \optpotentialunscaledv||_{\laplacianglobal} \leq \delta ||\optpotentialunscaledv||_{\laplacianglobal}
\end{align}
In time $\tilde{O}(m k^2 \sqrt{\kappa} \poly(\epsilon^{-1}))$.
\end{lemma}

\Proof
Applying Lemma \ref{lem:laplaciansolver} to each of the Laplacians that make up $\tilde{\laplacianglobal}$,
we can obtain a linear operator $A$ such that:
\begin{align}
\tilde{\laplacianglobal}^{+} \preceq A \preceq 2\tilde{\laplacianglobal}^{+}
\end{align}

Such that $A \vecx$ can be evaluated in time $\tilde{O}(mk)$.

Combining these bounds then gives:
\begin{align}
A
\preceq 2 \tilde{\laplacianglobal} ^{+}
\preceq 2 \kappa A
\end{align}

Then the running time follows from Lemma \ref{lem:preconcheby}, which requires an extra
$\kappa$ iterations, and the fact that a forward multiply involving $\pd$ costs $O(mk^2)$.
\QED

Using this extension to the solver we can prove our main theorem about solving
quadratically coupled flows, which we prove in Appendix \ref{sec:solvererror}.

\begin{theorem}
\label{thm:approxelectrical}
There is an algorithm \textsc{QuadraticallyCoupledFlow}
such that for any $\delta > 0$ and $F > 0$, any set of energy matrices $\pd(e)$ such that
$I \preceq \pd(e) \preceq \maxratio I$ for a parameter $\maxratio$
and $\kappa \lambda_{\min}(\pd(e)) \geq \lambda_{\max}(\pd(e))$, and any
demand vector $\demandv$ such that the corresponding minimum energy flow is $\optflowv$,
computes in time $\tilde{O}((\sqrt{\kappa}k^2 + k^\omega) m \log(\maxratio /\delta))$
a vector of vertex potentials $\approxpotentialvertv$ and a flow $\approxflowv$ such that

\begin{enumerate}
\item
\label{part:conservationofflow}
$\approxflowv$ satisfies the demands in all the commodities
$\edgevertexglobal^T \approxflowv = \demandv$
\item
\label{part:flowenergy}
$
\eflow_{\approxflowv}(\pd) \leq (1+\delta) \eflow_{\optflowv}(\pd)
$
\item
\label{part:flowdifference}
for every edge $e$,
$
|\power_{\optflow}(\pd, e)  - \power_{\approxflow}(\pd, e)|
\leq \delta \eflow_{\optflow}(\pd)
$
\item
\label{part:potentialobjective}
The energy given by the potentials $\epotential(\pd, \approxpotentialvertv)$ is at most $1$.
and its objective, $\demandv^T \approxpotentialvertv$ is at least
$(1 - \delta) \econductance(\pd)$.
\end{enumerate}
\end{theorem}

\section{Approximately Solving Quadratically Capacitated Flows}
\label{sec:inner}

We now show that we can repeatedly solve quadratically coupled flows
inside a multiplicative weights routine to minimize
the maximum saturation of an edge.
Pseudocode of our algorithm is shown in Algorithm \ref{alg:multiplicativeweights}.

\begin{algo}[h]
\qquad

\textsc{QuadraticallyCapacitatedFlow}
\vspace{0.05cm}

\underline{Input:} Weighted graph $G=(V,E,w)$,
energy matrix $\pd(e)$ for each edge $e$,
demands $\demandv_1 \ldots \demandv_k$ for each commodity.
Error bound $\epsilon$.

\underline{Output:} Either a collection of flows $\flowv$
such that $\saturation{\flowv}{\pd}{e} \leq 1 + \epsilon$,
or FAIL indicating that there does not exist a solution
$\flowv$ where $\saturation{\flowv}{\pd}{e} \leq 1 - 2 \epsilon$.

\vspace{0.2cm}

\begin{algorithmic}[1]
\STATE{$\rho \leftarrow 10 m^{1/3} \epsilon^{-2/3}$}
\STATE{$\numiter \leftarrow 20 \rho \ln{m} \epsilon^{-2} = 200 m^{1/3} \ln{m} \epsilon^{-8/3}$}
\STATE{Initialize $\weight^{(0)}(e) = 1$ for all $e \in E$}
\STATE{$\flowv \leftarrow \zerosv$}
\STATE{$\numiter_1 \leftarrow 0$}
\FOR{$t = 1 \ldots \numiter$}
    \STATE{Compute $\mu^{(t-1)} = \sum_{e} \weight^{(t - 1)}(e)$}
    \STATE{Compute reweighed energy matrices,
        $\pd^{(t-1)}(e) = \left( w^{(t-1)}(e) + \frac{\epsilon}{m} \mu^{(t - 1)} \right) \pd(e)$} \label{line:iterationweights}
	\STATE{Query $\textsc{QuadraticallyCoupledFlow}$ with energy matrices $\pd^{(t - 1)}(e)$
        and error bound $\delta = \frac{\epsilon}{m}$, let the flow returned be $\approxflowv^{(t)}$}
	\IF{$\eflow_{\approxflowv^{(t)}}(\pd^{(t - 1)}) > \mu^{(t-1)}$}
		\RETURN{\fail}
	\ELSE
        \IF{$\saturation{\approxflowv^{(t)}}{\pd^{(t - 1)}}{e} \leq \rho$ for all $e$}
            \STATE{$\flowv \leftarrow \flowv + \approxflowv^{(t)}$}
            \STATE{$\numiter_1 \leftarrow \numiter_1 + 1$}
        \ENDIF
		\FOR{$e \in E$}
			  \STATE{$\weight^{(t)}(e) \leftarrow \weight^{(t-1)}(e) \left(1+\frac{\epsilon}{\rho}
                \saturation{\approxflowv^{(t)}}{\pd^{(t - 1)}}{e} \right) $}
		\ENDFOR
	\ENDIF
\ENDFOR
\RETURN {$\frac{1}{\numiter_1} \flowv$}
\end{algorithmic}

\caption{Multiplicative weights update routine for approximately solving
quadratically capacitated flows}\label{alg:multiplicativeweights}
\end{algo}

The guarantees of this algorithm can be formalized as follows:

\begin{theorem}
\label{thm:multiplicativeweights}
Given a graph $G = (V, E)$ and energy matrices $\pd(e)$ on each of the edges
such that $\kappa \lambda_{min} \pd(e) > \lambda_{max} \pd(e)$ and a parameter
$\epsilon$, \textsc{QuadraticallyCapacitatedFlow} returns one of the
following in $\tilde{O}(mk^2\kappa \epsilon^{-8/3})$ time:

\begin{itemize}
\item A $k$-commodity flow $\approxflowv$ such that:
\[
\saturation{\flowv}{\pd}{e} \leq 1 + 10\epsilon
\]
for all edges $e$ and
\[
\edgevertexglobal \flowv = \demandv
\]

\item $\fail$ indicating that there does not exist a $k$-commodity flow $\flowv$
that satisfies all demands and have
\[
\saturation{\flowv}{\pd}{e} \leq 1 - \epsilon
\]
on all edges.
\end{itemize}
\end{theorem}

We first state the following bounds regarding the overall sum of potentials $\mu^{(t)}$,
the weight of a single edge $\weight^{(t)}(e)$ and the effective conductance given by
the reweighed energy matrices at each iteration, $\epotential (\pd^{(t)})$.

\begin{lemma}
\label{lem:potentials}
The following holds when $\approxflowv$ satisfies
\[
\sum_e \saturation{\approxflowv}{\pd^{(t-1)}}{e} \leq \mu^{(t-1)}
\]

\begin{enumerate}
    \item \label{part:muupper}
        \begin{align}
             \mu^{(t)} \leq \exp \left( \frac{\epsilon}{\rho} \right) \mu^{(t-1)}
        \end{align}
    \item \label{part:weightlower}
        $\weight^{(t)}(e)$ is non-decreasing in all iterations, and if 		 $\saturation{\approxflowv^{(t)}}{\pd^{(t)}}{e} \leq \rho$,
		we have:
            \begin{align}
                \weight^{(t)}(e) \geq \exp \left(\frac{\epsilon}{\rho} \saturation{\approxflowv^{(t)}}{\pd^{(t)}}{e} \right) \weight^{(t-1)}(e)
            \end{align}
    \item \label{part:potentialincrease}
        If for some edge $e$ we have $\saturation{\approxflowv^{(t)}}{\pd}{e} \geq \rho$, then
            $\econductance(\pd^{(t)}) \geq \econductance(\pd^{(t-1)}) \exp \left( \frac{\epsilon^2 \rho^2}{5m} \right)$
\end{enumerate}
\end{lemma}

The proof of Lemma \ref{lem:potentials} relies on the following facts about $\exp(x)$ when $x$ is close to 1:

\begin{fact}
\label{fact:log}
\begin{enumerate}
   \item \label{part:logupper}
   If $x \geq 0$, $1 + x \leq \exp(x)$.
   \item \label{part:loglower}
   If $0 \leq x \leq \epsilon$, then $1 + x \geq \exp((1-\epsilon)x)$.
\end{enumerate}
\end{fact}

\Proofof{Part \ref{part:muupper}}
\begin{align}
\mu^{(t)}
= & \sum_e \weight^{(t)}(e) \nonumber \\
= & \sum_e \weight^{(t-1)}(e) (1+\frac{\epsilon}{\rho} \saturation{\approxflowv^{(t)}}{\pd}{e})
\qquad \text{By the update rule} \nonumber \\
= & \left( \sum_e \weight^{(t-1)}(e) \right)
    + \frac{\epsilon}{\rho} \left( \sum_e \weight^{(t-1)}(e) \saturation{\approxflowv^{(t)}}{\pd}{e} \right) \nonumber \\
\leq & \mu^{(t-1)} + \frac{\epsilon}{\rho} \mu^{(t-1)}
\qquad \text{By definition of $\mu^{(t-1)}$ and total weighted saturation} \nonumber \\
= & (1 + \frac{\epsilon}{\rho}) \mu^{(t-1)}
\leq \exp(\frac{\epsilon}{\rho}) \mu^{(t-1)}
\qquad \text {By Fact \ref{fact:log}.\ref{part:logupper}}
\end{align}

\QEDpart{Part \ref{part:muupper}}

\Proofof{Part \ref{part:weightlower}}

If $\saturation{\approxflowv^{(t)}}{\pd^{(t)}}{e} \leq \rho$, then $\frac{\epsilon}{\rho} \saturation{\approxflowv^{(t)}}{\pd^{(t)}}{e} \leq \epsilon$
and:

\begin{align}
\weight^{(t)}(e)
= & \weight^{(t-1)}(e) \left( 1 + \frac{\epsilon}{\rho} \saturation{\approxflowv^{(t)}}{\pd^{(t)}}{e} \right) \nonumber \\
\leq & \weight^{(t-1)}(e) \exp \left( \frac{\epsilon (1-\epsilon) }{\rho} \saturation{\approxflowv}{\pd}{e} \right)
\qquad \mbox{By Fact \ref{fact:log}.\ref{part:loglower}}
\end{align}

\QEDpart{Part \ref{part:weightlower}}

\Proofof{Part \ref{part:potentialincrease}}

Let $e$ be the edge where $\saturation{\approxflowv}{\pd}{e} \geq \rho$, then
since $\pd^{(t-1)}(e) \succeq \frac{\epsilon}{m} \mu I$ by line \ref{line:iterationweights},
we have:

\begin{align}
\saturation{\approxflowv}{\pd^{(t-1)}}{e}^2
\geq & \frac{\epsilon}{3m} \mu^{(t-1)} \rho^2 \nonumber \\
\geq & \frac{\epsilon \rho^2}{3m} \eflow(\pd^{(t-1)}, \approxflowv)
\qquad \mbox{By assumption of the energy of the flow returned} \label{eq:energyfraction}
\end{align}

Invoking the guarantees proven in Theorem \ref{thm:approxelectrical},
we have:

\begin{align}
\saturation{\optflowv}{\pd^{(t-1)}}{e}^2
\geq & \saturation{\approxflowv}{\pd^{(t-1)}}{e}^2
    -   |\saturation{\optflowv}{\pd^{(t-1)}}{e}^2
    - \saturation{\approxflowv}{\pd^{(t-1)}}{e}^2| \nonumber \\
\geq & \saturation{\approxflowv}{\pd^{(t-1)}}{e}^2  - \delta  \eflow(\pd^{(t-1)})
\qquad \mbox{By Part \ref{part:flowdifference}} \nonumber\\
\geq & \frac{\epsilon \rho^2}{3m} \eflow_{\approxflowv^{(t)}}(\pd^{(t-1)})
	-  \delta  \eflow(\pd^{(t-1)})
\qquad \mbox{By Equation \ref{eq:energyfraction}} \nonumber \\
\geq & \frac{\epsilon \rho^2}{3(1+\delta)m} \eflow(\pd^{(t-1)}) - \delta \eflow(\pd^{(t-1)})
\qquad \mbox{By Part \ref{part:flowenergy}} \nonumber \\
\geq & \frac{\epsilon \rho^2}{4m} \eflow(\pd^{(t-1)})
\end{align}

Then by the relation between $\optpotentialvertv$ and $\optflowv$, we have that
\begin{align}
    \burn_{\optpotentialvertv^{(t - 1)}}(\pd^{(t-1)}(e))
    \geq & \frac{\epsilon \rho^2}{4m} \epotential_{\optpotentialvertv^{(t - 1)}}(\pd^{(t-1)})
\label{eq:erfraction}
\end{align}

Then since $\weight^{(t)}(e) \geq (1+\epsilon) \weight^{(t - 1)}(e)$, using the current
set of optimal potential gives:

\begin{align}
\epotential_{\optpotentialvertv^{(t - 1)}}(\pd^{(t)})
\leq & (1 - \frac{\epsilon^2 \rho^2}{4m}) \epotential(\pd^{(t-1)}, \optpotentialvertv) \label{eq:newer}
\end{align}

Which means that when $\epsilon < 0.01$,
$\sqrt{1 + \frac{\epsilon^2 \rho^2}{5m}} \optpotentialvertv^{(t-1)}$ is a valid set of
potentials for $\pd^{(t)}$ and therefore:

\begin{align}
\econductance(\pd^{(t)})
\geq & (\demandv^T \sqrt{1 + \frac{\epsilon^2 \rho^2}{5m}} \optpotentialvertv^{(t-1)})^2 \nonumber \\
\geq & \exp \left( \frac{\epsilon^2 \rho^2}{5m} \right) (\demandv^T \optpotentialvertv^{(t-1)})^2 \nonumber \\
= & \exp \left( \frac{\epsilon^2 \rho^2}{5m} \right) \econductance(\pd^{(t-1)})
\end{align}

\QEDpart{Part \ref{part:potentialincrease}}

\Proofof{Theorem \ref{thm:multiplicativeweights}}

Since $\sum_e (w^{(t-1)}(e) + \frac{\epsilon}{m} \mu^{(t-1)})  = (1+ \epsilon)\mu^{(t-1)}$,
if there exist a flow $\flowv$ such that $\saturation{\flowv}{\pd}{e} \leq 1 - 2 \epsilon$
for all $e$, we have that $\eflow_{\optflowv^{(t)}}(\tilde{\pd}) \leq (1 - \epsilon) \mu^{(t-1)}$.
Then if the algorithm does not return $\fail$, Theorem \ref{thm:approxelectrical}
means that $\approxflowv^{(t)}$ satisfies:

\begin{align}
\eflow_{\approxflowv^{(t)}}
\leq & (1+\delta) (1 - \epsilon) \mu^{(t-1)} \nonumber \\
\leq & \mu^{(t-1)} \\
\sum_{e} \weight^{(t-1)}(e) \saturation{\approxflowv^{(t)}}{\pd^{(t-1)}}{e}^2
    \leq & \sum_{e} \weight^{(t-1)}(e)
\end{align}

Multiplying both sides by $\mu^{(t-1)}$ and
applying the Cauchy-Schwarz inequality gives:

\begin{align}
\left( \sum_{e} \weight^{(t-1)} (e) \right)^2
\geq & \left( \sum_{e} \weight^{(t-1)} (e) \right)
    \left( \sum_{e} \weight^{(t-1)} (e) \saturation{\approxflowv^{(t)}}{\pd^{(t-1)}}{e}^2 \right) \nonumber \\
\geq & \left( \sum_{e} \weight^{(t-1)} (e) \saturation{\approxflowv^{(t)}}{\pd^{(t-1)}}{e} \right)
\end{align}

Taking the square root of both sides gives:

\begin{align}
\sum_{e} \weight^{(t-1)}(e) \saturation{\approxflowv^{(t)}}{\pd^{(t-1)}}{e}
\leq & \mu^{(t-1)}
\end{align}

Therefore inductively applying Lemma \ref{lem:potentials} Part\ref{part:muupper},
we have:

\begin{align}
\mu^{(\numiter)}
\leq & \mu^{(0)} \cdot \left( \exp (\frac{\epsilon}{\rho}) \right)^\numiter \nonumber \\
= & \exp \left( \frac{\epsilon \numiter}{\rho}\right) m \nonumber \\
\leq & \exp \left( \frac{21 \ln m}{\epsilon} \right)
\end{align}

We now bound $N'$, the number of iterations $t$ where there is an edge with
$\saturation{\approxflowv^{(t)}}{\pd^{(t-1)}}{e} \geq \rho$.
Suppose $\econductance(\pd^{(0)}) \leq 1/2$, then in the flow returned,
no edge $e$ has $\saturation{\approxflowv^{(0)}}{\pd^{(0)}}{e} \geq 1$, which means
that the algorithm can already return that flow.

Then by the monotonicity of $\econductance(\pd^{(t)})$ and Lemma \ref{lem:potentials}
Part \ref{part:potentialincrease}, we have:

\begin{align}
\econductance(\pd^{(\numiter)})
\geq & 1/2 \cdot \exp \left( \frac{\epsilon^2 \rho^2 \numiter'}{5m} \right)
\end{align}

Combining this with
$\econductance(\pd^{(t)})
\leq \mu^{(\numiter)}$
gives:

\begin{align}
\frac{\epsilon^2 \rho^2 \numiter'}{5m}
\leq & \frac{21 \ln m}{\epsilon} \nonumber \\
\numiter'
\leq & \frac{105 m \ln m}{\rho^2  \epsilon^3} \nonumber \\
\leq & \epsilon {\numiter}
\end{align}

Then in all the $\numiter - \numiter' \geq (1-\epsilon)\numiter$ iterations, we have
$\saturation{\approxflowv^{(t)}}{\pd^{(t - 1)}}{e} \leq \rho$ for all edges $e$.
Then we have:

\begin{align}
\saturation{\sum_{t} \approxflowv^{(t)}}{\pd}{e}
\leq & \sum_{t} \saturation{\approxflowv^{(t)}}{\pd}{e}
	\qquad \mbox{Since $\pd(e)$ defines a norm} \nonumber \\
\leq & \log(\mu^{(t)}) / ( \frac{\epsilon}{\rho} )
	\qquad \mbox{By Lemma \ref{lem:potentials} Part \ref{part:weightlower}} \nonumber \\
= & \frac{1}{1 - \epsilon} T' \leq (1 + 2\epsilon) T'
\end{align}

\QED

\section{Algorithm for Maximum Concurrent Multicommodity Flow}
\label{sec:outer}

One of the main difficulties in directly applying the flow algorithm
from \cite{ChristianoKMST10} is that single commodity congestion constraints
of the form $||\flowv(e)||_1\leq u_e$ are 'sharper' than the $L_2$
energy functions due to the sign changes when each of the commodities
are around $0$.

As a result, we use the primal Primal-Dual SDP algorithm from \cite{AK07}
to generate the energy matrix.
Pseudocode of the outermost layer of our algorithm for maximum concurrent
flow is shown in Algorithm \ref{alg:outer}.

\begin{algo}[h]\label{alg:outer}
\qquad

\textsc{MaxConcurrentFlow}
\vspace{0.05cm}

\underline{Input:}
Capacitated graph $G = (V, E, \capacity)$ and demands $\demandv$.

Algorithm for computing energy matrices based on a list of $k$
flows, $\textsc{Energy}$ and for minimizing the maximum energy
along an edge.
Iteration count $N$ and error tolerance $\epsilon$.

\underline{Output:} Either a flow $\flowv$
that meets the demands and $||\flowv(e)||_1 \leq (1 + 10 \epsilon) \mu(e)$ on
all edges, or \fail indicating that there does not exist a flow
$\flowv$ that meets the demands and satisfy
$||\flowv(e)||_1 \leq \mu(e)$ on all edges.

\vspace{0.2cm}

\begin{algorithmic}[1]
\STATE{$\rho \leftarrow \sqrt{k \epsilon^{-1}}$}
\STATE{$\epsilon_1 \leftarrow \frac{\epsilon}{k \rho} = \frac{\epsilon^{1/2}}{k^{3/2}} $}
\STATE{$\epsilon_1' \leftarrow -\ln(1 - \epsilon_1)$}
\STATE{$N \leftarrow \rho \epsilon_1'^{-2} \log{k} = k^{7/2} \epsilon^{-3/2} \log{k}$}
\STATE{Initialize $\matsumm^{0}(e) = \zerosm$, $\matw^{0}(e) = \identity$}
\FOR{$t = 1 \ldots N$}
    \FOR{$e \in E$}
        \STATE{$\pd^{(t - 1)}(e) = \frac{1}{\capacity^2} \left( \matw^{(t-1)}(e) / ||\matw^{(t-1)}||_\infty
            + \epsilon \identity \right)$}
    \ENDFOR
    \STATE{Query \textsc{QuadraticallyCapacitatedFlow} with matrix $\pd^{(t - 1)}$}
	\IF{\textsc{QuadraticallyCapacitatedFlow} returns \fail}
		\RETURN{\fail}
    \ELSE
        \STATE{Let the flow returned be $\approxflowv^{(t)}$}
        \FOR{$e \in E$}
            \STATE{$(\matsumm^{(t)}(e), \matw^{(t)}(e)) \leftarrow
			\textsc{Update}(\matsumm^{(t - 1)}(e), \matw^{(t - 1)}(e), \approxflowv^{(t)})$}
        \ENDFOR
	\ENDIF
\ENDFOR
\RETURN {$\frac{1}{N} \sum_{t = 1}^N \approxflowv^{(t)}$}
\end{algorithmic}

\caption{Algorithm for minimizing $L_1$ congestion}\label{alg:outer}
\end{algo}

Where the update routine, $\textsc{UPDATE}$ is shown in Algorithm \ref{alg:update}.
Note that $\indicator_i$ indicates the matrix that's $1$ in entry $(i, i)$
and $0$ everywhere else, and $\matsumm^{(t)}$ is used to store the sum
of $\matm^{(r)}$ over $1 \leq r \leq t$ to we do not need to pass all of
them to each invocation of $\textsc{Update}$.

\begin{algo}[h]
\qquad

\textsc{Update}
\vspace{0.05cm}

\underline{Input:}
$\matsumm^{(t - 1)}(e), \matw^{(t - 1)}(e)$ from previous iterations,
flow from iteration $t$, $\approxflowv^{(t)}(e)$,
Capacity $\capacity(e)$,
Parameter $\epsilon_1'$.

\underline{Output:}
Sum matrix from current iteration $S$,
Energy matrix $X$.
\vspace{0.2cm}

\begin{algorithmic}[1]
\STATE{Find $i^{(t - 1)}(e) = \arg\max_i \matw^{(t-1)}_{ii}(e)$}
\STATE{$\matm^{(t)}(e) \leftarrow
    \frac{1}{2\rho} \left( (1+2\epsilon) \indicator_{i^{(t-1)}(e)}
	 -  \frac{1}{\capacity(e)^2} \flowv^{(t)}(e) (\flowv^{(t)}(e))^T + \rho \identity \right)$ \\
    Where $\indicator_i$ has $1$ in $(i,i)$ and $0$ everywhere else}
\STATE{$\matsumm^{(t)}(e) \leftarrow \matsumm^{(t - 1)}(e) + \matm^{(t)}(e)$}
\STATE{$\matw^{(t)}(e) = \exp(-\epsilon_1' \matsumm^{(t)}(e))$} \label{ln:assignw}
\RETURN{$(\matsumm^{(t)}(e), \matw^{(t)}(e))$}
\end{algorithmic}

\caption{Matrix exponential based algorithm for generating energy matrix}
\label{alg:update}
\end{algo}

We start off by bounding the condition number of $\pd^{(t)}(e)$.

\begin{lemma}
\label{lem:conditionnumber1}
\begin{align*}
\lambda_{max}(\pd^{(t)}(e)) \leq 2 k \epsilon^{-1} \lambda_{min}(\pd^{(t)}(e))
\end{align*}
\end{lemma}

\Proof
Since $\matw(e)$ is positive semi definite, we have $\pd^{(t)}(e) \succeq \epsilon \identity$.
Also, by definition of $i^{(t)}(e)$, we have that the maximum diagonal entry of
$\matw^{(t-1)}(e) / ||\matw^{(t-1)}(e)||_\infty$ is $1$.
Therefore $\trace(\pd^{(t)}(e)) \leq k + \trace(\epsilon \identity) \leq 2k$.
This in turn implies $\lambda_{max}(\pd^{(t)}(e)) \leq 2k$, which gives the required result.
\QED

We first show that when a flow exist,
\textsc{QuadraticallyCapacitatedFlow} returns a flow $\approxflowv^{(t)}(e)$
with low energy on each edge.

\begin{lemma}
\label{lem:fexist}

If there is $\optflowv$ such that $|\optflowv(e)|_1 \leq 1$,
then at each iteration $t$ we have
$\eflow_{\approxflow^{(t)}} (\pd^{(t-1)}, e) \leq 1 + 2 \epsilon$.
\end{lemma}

\Proof

\begin{align}
\eflow_{\optflowv(e)}(\matw^{(t-1)}(e))
= & \optflowv(e)^T \matw^{(t-1)(e)} \optflowv \nonumber \\
= & \sum_{ij} \optflowv_i(e) \optflowv_j(e) \matw^{(t-1)}_{ij}(e) \nonumber \\
\leq & \sum_{ij} |\optflow_i(e)| |\optflowv_j(e)| |\matw^{(t-1)}_{ij}(e)| \nonumber \\
\leq & ||\matw^{(t-1)}||_\infty ( \sum_{ij} |\optflowv_i(e)| |\optflowv_j(e)|  ) \nonumber \\
= & ||\matw^{(t-1)}||_\infty \left( \sum_{i} |\optflowv_i(e)| \right)^2 \nonumber \\
\leq & ||\matw^{(t-1)}(e)||_\infty \capacity(e)^2
\end{align}

Therefore we have:

\begin{align}
\eflow_{\approxflow^{(t)}} (\pd^{(t-1)}, e)
\leq & \frac{1}{\capacity(e)^2} \left(
    \frac{1}{||\matw^{(t-1)}(e)||_\infty} \eflow_{\optflowv(e)}(\matw^{(t-1)(e)})
        + \eflow_{\optflowv(e)}(\epsilon \identity)) \right) \nonumber \\
\leq & 1 + 2 \epsilon
\end{align}

Since $\eflow_{\optflowv(e)}(\epsilon \identity))
= \epsilon ||\optflowv(e)||_2^2
\leq \epsilon ||\optflowv(e)||_1^2
\leq \capacity(e)^2$.

And the properties of $\flowv^{(t)}(e)$ follows from the guarantees
of Theorem \ref{thm:multiplicativeweights}.
\QED

Using this width bound, we can now
 adapt the analysis in \cite{AK07} to show that the sum of flows
can be bounded in the matrix sense.

\begin{lemma}
\label{lem:diagheavy}
If in all iterations the flows returned satisfy $\flowv^{(t)}(e) (\flowv^{(t)}(e))^T \preceq \rho \capacity^2 I$,
then we have the following by the end of $N$ iterations:

\begin{align*}
\left( \sum_{t = 1}^N  \frac{1}{\capacity(e)^2} \flowv^{(t)}(e)(\flowv^{(t)}(e))^T \right)
\preceq & (1 + 2 \epsilon) \capacity^2 \sum_{t = 0}^{N - 1} \indicator_{i^{(t)}}
    + \frac{\epsilon N}{k} \identity
\end{align*}


\end{lemma}

\Proof

The proof is similar to the proofs of Theorems 10 and 1
in Section 6 of \cite{AK07}.
We have:

\begin{align}
\trace \left( \matw^{(t)}(e) \right)
= & \trace \left( \exp(-\epsilon'_1 \sum_{r=0}^t \matm^{(r)}(e) ) \right) \nonumber \\
\leq & \trace \left( \exp(-\epsilon'_1 \sum_{r=0}^{(t-1)} \matm^{(r)}(e) ) \exp(-\epsilon'_1 \matm^{(t)}(e) ) \right)
	\qquad\mbox{By the Golden-Thompson inequality} \nonumber \\
= & \trace \left( \matw^{(t-1)}(e) \exp( -\epsilon'_1 \matm^{(t)}(e) ) \right)\nonumber \\
\leq & \trace \left( \matw^{(t-1)}(e) (\identity -\epsilon_1 \matm^{(t)}(e) ) \right)
	\qquad \mbox{Since $\exp(-\epsilon'_1 \mata) \preceq (\identity - \epsilon'_1 \mata)$ when $0 \preceq A \preceq \identity$}\nonumber \\
= & \trace(\matw^{(t-1)}(e)) - \epsilon_1 \trace( \matw^{(t-1)}(e) \matm^{(t)} )
\end{align}

The construction of $\matw(e)$ from line \ref{ln:assignw} of \textsc{UPDATE}
means that $\matw(e)$ is positive semi-definite.
This in turn implies that $||\matw(e)||\infty \leq \max_{i} \matw_{ii}$.
Substituting gives:

\begin{align}
\trace\left( \matw^{(t-1)}(e) \matm^{(t)}(e) \right)
= & \trace \left( \matw^{(t-1)(e)} \cdot \left( (1 + 2 \epsilon) \indicator_{i^{(t-1)}}
    - \flowv^{(t)(e)} (\flowv^{(t)}(e))^T + \rho \identity \right) / 2 \rho \right) \nonumber \\
= & \frac{1}{2 \rho} \left( (1 + 2 \epsilon) ||\matw^{(t-1)}(e)||_\infty
	- \frac{1}{\capacity(e)^2} (\flowv^{(t)}(e))^T \matw^{(t-1)}(e) \flowv^{(t)}(e) \right)
	+ \frac{1}{2} \trace(\matw^{(t-1)}(e) )\nonumber \\
\leq & \frac{1}{2 \rho} \left( (1 + 2 \epsilon) ||\matw^{(t-1)}(e)||_\infty
	- ||\matw^{(t-1)}(e)||_\infty (\flowv^{(t)}(e))^T \pd^{(t-1)}(e) \flowv^{(t)}(e) \right)       \nonumber \\ 
&  \qquad + \frac{1}{2} \trace(\matw^{(t-1)}(e))
\qquad \mbox{Since $\frac{\matw^{(t-1)}(e)}{\capacity(e)^2 ||\matw^{(t-1)}(e)||_\infty} \preceq \pd^{(t-1)}(e)$} \nonumber \\
\geq & \frac{1}{2} \trace(\matw^{(t-1)}(e))
    \qquad \mbox{By Lemma \ref{lem:fexist}}
\end{align}

Combining these two gives:

\begin{align}
\trace \left( \exp(-\epsilon_1' \matsumm^{(N)}(e)) \right)
= & \trace \left( \matw^{N}(e) \right) \nonumber \\
\leq & k \left(1 - \frac{\epsilon_1}{2} \right)^N \nonumber \\
\leq & k \exp\left( - \frac{\epsilon_1}{2} N \right) \nonumber \\
\leq & \exp\left( - \frac{(1 - \epsilon_1)\epsilon_1}{2} N \right)
	\qquad \mbox {Since $ N = 10 \rho \epsilon_1'^{-2} \log{k}$}
\end{align}

Using the fact that $\exp(- \lambda_{\max}(\mata)) \leq \trace(- \exp(\mata))$, we get:

\begin{align}
\frac{(1 - \epsilon_1)\epsilon_1}{2} \numiter \identity
\preceq & \epsilon_1' \sum_{t=0}^{\numiter} \matm^{(t)}(e) \nonumber \\
= & \epsilon_1' \sum_{t=0}^{\numiter} \left( (1 + 2 \epsilon)
    \indicator_{i^{(t)}} - \frac{1}{\capacity^2} \flowv^{(t)}(e)(\flowv^{(t)}(e))^T + \rho \identity \right) / 2 \rho \\
- \epsilon_1' \rho N \identity
\preceq & \sum_{t = 0}^{\numiter} \left( (1 + 2 \epsilon)
    \indicator_{i^{(t)}} - \frac{1}{\capacity^2} \flowv^{(t)}(e)(\flowv^{(t)}(e))^T \right) \\
\frac{1}{\capacity^2} \left( \sum_{t = 1}^{\numiter}  \flowv^{(t)}(e)(\flowv^{(t)}(e))^T \right)
\preceq & (1 + 2\epsilon) \sum_{t = 0}^{\numiter - 1} \indicator_{i^{(t)}}
    + \epsilon_1' \rho N \identity
\end{align}

Substituting in the setting of $\epsilon_1 = \frac{\epsilon}{k \rho}$ gives the desired result.

\QED

This in turns lets us bound the $L_1$ congestion of the flow returned
after $T$ iterations.

\begin{theorem}
\label{thm:outermatrixexp}
After $N = \tilde{O}(k^{7/2} \epsilon^{-5/2} \log{k})$ iterations,
\textsc{MaxConcurrentFlow} returns a flow $\flowv$ where for each edge $e$, we have:

$$\left|\left| \frac{1}{\numiter} \sum_{t = 1}^{\numiter} \flowv^{(t)} \right|\right|_1
\leq (1 + 3\epsilon) \capacity$$


\end{theorem}

\Proof

We first bound the width of each update step.
Note that by construction we have:

\begin{align}
\frac{\epsilon}{\capacity^2 k} (\flowv^{(t)}(e))^T \identity \flowv^{(t)}(e)
    \leq & (1+\epsilon) \nonumber \\
\frac{1}{\capacity^2}(\flowv^{(t)}(e))^T \identity \flowv^{(t)}(e)
    \leq & 2k\epsilon^{-1}
\end{align}

Then by the Cauchy-Schwarz inequality we have:

\begin{align}
\left|\left| \frac{\flowv^{(t)}(e)}{\capacity(e)} \right|\right| _1^2
\leq & k \left|\left| \frac{\flowv^{(t)}(e)}{\capacity(e)} \right|\right| _2^2 \nonumber \\
= & 2k\epsilon^{-1}
\end{align}

Which gives that $\rho = \sqrt{2k\epsilon^{-1}}$ suffices as width parameter.

Let $\signv$ be the vector corresponding to the signs of the entries of
$\sum_{t = 1}^{\numiter} \flowv^{(t)}(e)$, aka.
$\signv^T \sum_{t = 1}^{\numiter} \flowv^{(t)}(e)
    = || \sum_{t = 1}^{\numiter} \flowv^{(t)}(e) ||_1$.
Then:

\begin{align}
\left|\left|\frac{1}{\capacity \numiter} \sum_{1 \leq t \leq \numiter} \flowv^{(t)}(e) \right|\right|^2_1
= & \left( \signv^T \frac{1}{\numiter} \sum_{t = 1}^N \flowv^{(t)}(e) \right)^2 \nonumber \\
= & \frac{1}{\capacity^2 \numiter^2} \left( \signv^T \sum_{t=1}^\numiter \flowv^{(t)}(e) \right)^2 \nonumber \\
\leq & \frac{1}{\numiter} \sum_{t=1}^\numiter \left( \signv^T \flowv^{(t)}(e) \right)^2
	\qquad \mbox {By the Cauchy-Schwarz inequality} \nonumber \\
= &\frac{1}{\numiter} \signv^T \left( \sum_{t=1}^\numiter \frac{1}{\capacity^2} \flowv^{(t)}(e)(\flowv^{(t)}(e))^T \right) \signv \nonumber \\
\leq & \frac{1}{\numiter} \signv^T
    \left( \sum_{t=1}^\numiter (1 + 2 \epsilon) \indicator_{i^{(t)}} + \frac{\epsilon \numiter}{k} \identity\right) \signv
	\qquad \mbox {By Lemma \ref{lem:diagheavy}} \nonumber \\
= & 1+3 \epsilon
~~~~~\text{Since $\sign_i = \pm 1$ and $\signv^T \signv = k$}
\end{align}

\QED

The running time of the algorithm can then be bounded as follows:

\begin{lemma}
Each iteration of the \textsc{MaxConcurrentFlow} runs in $\tilde{O}(m^{4/3}k^{5/2}\epsilon^{-19/6} + mk^{\omega})$ time,
giving an overall running time of $\tilde{O}(m^{4/3}k^{6}\epsilon^{-17/3} + mk^{7/2+\omega}\epsilon^{-5/2})$, where $\omega$
is the matrix multiplication exponent.
\end{lemma}

\Proof
The first term follows from the running time of \textsc{QuadraticallyCapacitatedFlow}
proven in Theorem \ref{thm:multiplicativeweights}
and $\kappa(\pd(e)) \leq O(\sqrt{k \epsilon^{-1}})$ from Lemma \ref{lem:conditionnumber1}.
For the second term, the bottleneck is the computation of matrix exponentials.
This can be done in $\poly(\log{k})$ matrix multiplies using \cite{YL93},
giving the $\tilde{O}(k^{\omega})$ bound in the each of the iterations.
\QED

\section{Alternative Outer Algorithm}
\label{sec:outer1}

We show a modified formulation of the capacity bounds
as $2^{k}$ constraints per edge that brings us back to
minimizing the maximum congestion.
This gives a more combinatorial approach to minimizing the maximum $L_1$
congestion, although the algorithm is slightly more intricate.
As the computation of energy matrices only rely on the sum of flows so far
(aka. history independent), we describe its computation in a separate routine
$\textsc{ENERGY}$ and first state the overall algorithm in Algorithm \ref{alg:outer1}.

\begin{algo}[h]\label{alg:outer1}
\qquad

\textsc{MaxConcurrentFlow1}
\vspace{0.05cm}

\underline{Input:}
Capacitated graph $G = (V, E, \capacity)$ and demands $\demandv$.

Algorithm for computing energy matrices based on a list of $k$
flows, $\textsc{Energy}$ and for minimizing the maximum energy
along an edge, $\textsc{QuadraticallyCapacitatedFlow}$.
Width parameter $\rho$, iteration count $N$ and error tolerance $\epsilon$.

\underline{Output:} Either a flow $\approxflowv$
that meets the demands and $||\approxflowv(e)||_1 \leq (1 + 10 \epsilon) \mu(e)$ on
all edges, or \fail indicating that there does not exist a flow
$\flowv$ that meets the demands and satisfy
$||\flowv(e)||_1 \leq \mu(e)$ on all edges.

\vspace{0.2cm}

\begin{algorithmic}[1]
\FOR{$t = 1 \ldots \numiter$}
    \FOR{$e \in E$}
        \STATE{$\pd^{(t)}(e) = \textsc{Energy}(\sum_{1 \leq r < t} \flowv^{(r)}(e))$}
    \ENDFOR
	\STATE{Query \textsc{QuadraticallyCapacitatedFlow} with matrix $\pd^{(t)}$}
	\IF{\textsc{\textsc{QuadraticallyCapacitatedFlow}} returns \fail}
		\RETURN{\fail}
	\ENDIF
\ENDFOR
\RETURN {$\frac{1}{\numiter} \sum_{t = 1}^\numiter \flowv^{(\numiter)}$}
\end{algorithmic}

\caption{Alternate Algorithm for Maximum Concurrent Multicommodity Flow}\label{alg:outer1}
\end{algo}

We start with the following observation that the maximum among the sums
given by all $2^{k}$ choices of signs to $\flowv_i(e)$ equals congestion.

\begin{observation}
\[
\sum_{i = 1}^k |\flowv_{i}(e)| =
\max_{\sign_1, \sign_2 \ldots \sign_k \in \{-1, 1\}^{k}} \sum_i \sign_i \flowv_{i}(e)
\]
\end{observation}

We let $\signlist$ to denote the set of all $2^k$ settings of signs.
This allows us to reformulate the constraint of $||\flowv(e)||_1 \leq \capacity(e)$
as:

\begin{align}
\sign^T \flowv(e) \leq \capacity(e) \qquad \forall \sign \in \signlist
\end{align}

This reduces the problem back to minimizing the maximum among all
$|\signlist| = 2^k$ dot products with $\flowv(e)$.
To solve this problem we can once again apply the multiplicative weights
framework.
We state the convergence result in a more general form:

\begin{theorem}
\label{thm:roughweights}
If for all flows $\flowv(e)$, $\textsc{Energy}(\flowv(e))$ returns a matrix

\begin{align}
    \pd(e) = \frac{1}{\sum_{\signv \in \signlist} \approxweight(\signv)}
         \sum_{\signv \in \signlist} \frac{\approxweight(e)}{\capacity(e)^2} \signv \signv^T
         + \frac{\epsilon}{\capacity(e)^2 } \identity
\end{align}

Where $\approxweight(\signv)$ satisfies

\begin{align}
    \exp \left( \frac{\epsilon \signv^T \flowv(e)}{\rho \capacity(e)} \right)
    \leq & \approxweight(\signv)
    \leq (1 + \epsilon) \exp \left( \frac{\epsilon \signv^T \flowv(e)}{\rho \capacity(e)} \right)
\end{align}

Then:

\begin{enumerate}
\item \label{part:conditionnumber} $\lambda_{max}(\pd(e)) \leq 2k\epsilon^{-1} \lambda_{min}(\pd(e))$.
\item \label{part:convergence}
If there exist a flow $\optflowv$ that meets all the demands and
have $||\optflowv(e)||_1 \leq (1 - 3\epsilon) \capacity(e)$,
$\textsc{MaxConcurrentFlow}$ with $\rho = k$ and $\numiter = \rho k \epsilon^{-2}$
returns a flow $\approxflowv$ that meets the demands and satisfy
$||\approxflowv(e)||_1 \leq (1 + 3\epsilon) \capacity(e)$ over all edges.
\end{enumerate}

\end{theorem}

\Proofof{Part \ref{part:conditionnumber}}

For $\lambda_{min}(\pd(e))$, we have $\pd(e) \succeq \frac{\epsilon}{\capacity(e)^2} I$
and the condition number bound follows from $\lambda_{min}(I) = 1$.

We can bound the maximum eigenvalue with the trance. Note that $\trace(\signv \signv^T) = k$
since the diagonal of $\signv \signv^T$ is all $1$.
This gives:

\begin{align}
\lambda_{max}(\pd(e))
\leq & \trace(\pd(e)) \nonumber \\
= &  \frac{1}{\sum_{\signv \in \signlist} \approxweight(\signv)}
         \sum_{\signv \in \signlist} \frac{\approxweight(e)}{\capacity(e)^2} \trace{\signv \signv^T}
         + \frac{\epsilon}{\capacity(e)^2 } \trace{\identity} \nonumber \\
= & \frac{1+\epsilon}{\capacity(e)^2} k
\end{align}

Therefore we get:
\begin{align}
\frac{\lambda_{\max}(\pd(e))}{\lambda_{\min}(\pd(e))}
\leq & \frac{1+\epsilon}{\capacity(e)^2} k \epsilon^{-1} \nonumber \\
\leq & 2k \epsilon^{-1}
\end{align}

\QED

\Proofof{Part \ref{part:convergence}}

We define the exact set of weights that $\approxweight(\signv)$ are trying to approximate:

\begin{align}
\weight(\signv)^{(t)} = \exp \left( \frac{\epsilon}{\rho} \frac{\signv^T \flowv^{(t)}}{\capacity(e)} \right)
\end{align}

Also, let $\weightsum^{(t)} = \sum_{\signv \in \signlist} \weight(\signv)^{(t)}$.
We have that at any iteration:

\begin{align}
\weightsum^{(t)}
\geq & \max_{\signv} \weight(\signv)^{(t)} \nonumber \\
= & \exp( \max_{\signv} \flowv^{(t)} ) \nonumber \\
= & \exp( ||\flowv^{(t)}||_1 )
\end{align}

Therefore it suffices to upper bound the value of $\mu^{(t)}$.
First note that if there is a flow $\optflowv$ such that
$||\optflow(e)||_1 \leq (1 - 3 \epsilon) \capacity(e)$,
then we have $\signv^T \optflow(e) \leq ||\optflow(e)||_1 \leq \capacity(e) $ for any $\signv$
Squaring this and taking sum over all $\signv \in \signlist$ gives:

\begin{align}
\optflow(e)^T \pd(e)^{(t - 1)} \optflow(e)
= & \optflow(e) \left( \frac{1}{\sum_{\signv \in \signlist} \approxweight(\signv)^{(t-1)}}
         \sum_{\signv \in \signlist} \frac{\approxweight(e)^{(t-1)}}{\capacity(e)^2} \signv \signv^T
         + \frac{\epsilon}{k \capacity(e)^2} \identity \right) \optflow(e) \nonumber \\
= &  \frac{1}{\sum_{\signv \in \signlist} \approxweight(\signv)^{(t - 1)}}
         \sum_{\signv \in \signlist} \approxweight(\signv)^{(t - 1)}
            \left|\left|\frac{\signv^T \optflow(e)}{\capacity(e)}\right|\right|^2
    + \epsilon \left|\left|\frac{\optflow(e)}{\capacity(e)}\right|\right|_2^2 \nonumber \\
\leq & \frac{1}{\sum_{\signv \in \signlist} \approxweight(\signv)^{(t - 1)}}
         (1 - 3\epsilon)^2 \sum_{\signv \in \signlist} \approxweight(\signv)^{(t - 1)}
         + \epsilon \nonumber \\
\leq & (1 - 2\epsilon)^2
\end{align}

This means that there exist a flow $\flowv$ such that:

\begin{align*}
\saturation{\flowv}{\pd^{(t-1)}}{e} \leq 1 - 2 \epsilon
\end{align*}

Therefore by Theorem \ref{thm:multiplicativeweights}, we have that
for all edges $e$:
\begin{align*}
\saturation{\approxflowv^{(t)}}{\pd^{(t - 1)}}{e} \leq 1 - \epsilon
\end{align*}
This has two consequences:

\begin{enumerate}
\item   \label{part:maxcongestion}
\begin{align}
\frac{\epsilon}{\capacity(e)^2} (\approxflowv^{(t)})^T \identity \approxflowv^{(t)}
\leq & 1 - \epsilon
    \qquad \mbox{Since $\pd(e) \succeq \frac{\epsilon}{\capacity(e)^2}  \identity$} \nonumber \\
\left|\left|\frac{\approxflowv^{(t)}}{\capacity(e)}\right|\right|_1^2
\leq & k \left|\left|\frac{\approxflowv^{(t)}}{\capacity(e)}\right|\right|_2^2
= \frac{k}{\epsilon}
	\qquad \mbox{By the Cauchy-Schwarz inequality}
\end{align}

Squaring both sides gives that $\frac{||\approxflowv^{(t)}||_1}{\capacity(e)} \leq \sqrt{k} \epsilon^{-1/2}$,
which allows us to bound the width of the multiplicative updates.

\item \label{part:weightedcongestion}
Expanding out the first term in the formulation of $\pd(e)$ gives:

\begin{align}
\sum_{\signv \in \signlist} \approxweight(\signv)^{(t - 1)} \left( \frac{\signv^T \approxweight(e)}{\capacity(e)} \right)^2
\leq & (1 -  \epsilon) \sum_{\signv \in \signlist} \approxweight(\signv)^{(t-1)}
\end{align}

Multiplying both sides by $\sum_{\signv \in \signlist} \approxweight(\signv)^{(t-1)}$ and applying Cauchy-Schwarz
inequality gives:

\begin{align}
\sum_{\signv \in \signlist} \approxweight(\signv)^{(t-1)} \left| \frac{\signv^T \approxweight(e)}{\capacity(e)} \right|
\leq & \sum_{\signv \in \signlist} \approxweight(\signv)^{(t-1)}
\end{align}

Combining with the fact that $\weight \leq \approxweight \leq (1 + \epsilon) \weight$ gives:

\begin{align}
\sum_{\signv \in \signlist} \weight (\signv)^{(t-1)}
    \left| \frac{\signv^T \approxweight(e)}{\capacity(e)} \right|
\leq & (1 + \epsilon) \sum_{\signv \in \signlist} \weight(\signv)^{(t-1)}
\end{align}

\end{enumerate}

Using Fact \ref{fact:log} we have:

\begin{align}
\weightsum^{(t - 1)}
= & \sum_{\signv \in \signlist} \weight (\signv)^{(t - 1)}
    \exp \left( \frac{\epsilon}{\rho} \frac{\signv^T \approxweight(e)}{\capacity(e)} \right) \nonumber \\
\leq & \sum_{\signv \in \signlist} \weight (\signv)^{(t - 1)}
    \exp \left( \frac{\epsilon}{\rho} \left| \frac{\signv^T \approxweight(e)}{\capacity(e)} \right|  \right)
    \qquad \mbox{Since $\exp(x)$ is monotonic and $x \leq |x|$} \nonumber \\
\leq & \sum_{\signv \in \signlist} \weight (\signv)^{(t - 1)}
    \frac{( 1 + 2 \epsilon) \epsilon}{\rho} \left| \frac{\signv^T \approxweight(e)}{\capacity(e)} \right|
    \qquad \mbox{By Fact \ref{fact:log} Part \ref{part:loglower} and
        $\left| \frac{\signv^T \approxweight(e)}{\capacity(e)} \right| \leq \rho$} \nonumber \\
\leq & \frac{\epsilon}{\rho} (1 + 2\epsilon) (1 + \epsilon) \sum_{\signv \in \signlist} \weight(\signv)^{(t - 1)}
    \nonumber \\
\leq & \frac{\epsilon(1+4\epsilon)}{\rho} \weightsum^{(t - 1)} \nonumber \\
\leq & \exp(\frac{\epsilon(1+4\epsilon)}{\rho})\weightsum^{(t - 1)}
	\qquad \mbox{By Fact \ref{fact:log} Part \ref{part:logupper}}
\end{align}

Applying this inductively along with the fact that $\weightsum^{(0)} = 2^k$ gives:

\begin{align}
\weightsum^{(t)}
\leq & 2^k \exp \left( \frac{\epsilon(1+4\epsilon)}{\rho} \right) ^t \nonumber \\
= & \exp \left( \frac{\epsilon(1+4\epsilon)t }{\rho} + k \right)
\end{align}

Substituting in $\numiter = \frac{\rho k}{\epsilon^2}$ gives:

\begin{align}
\weightsum^{\numiter}
\leq & \exp \left( \frac{\epsilon(1+4\epsilon) \rho k }{\rho \epsilon^2} + k \right) \nonumber \\
= & \exp \left( \frac{(1+5\epsilon)k}{\epsilon} \right) \nonumber \\
= & \exp \left( \frac{\epsilon}{\rho}(1+5\epsilon \right) \numiter)
\end{align}

Which gives $\frac{1}{\numiter} |\flowv^{(\numiter)}(e)|_1 \leq (1+\epsilon) \capacity(e)$.

\QED

\subsection{Efficient Estimation of the Energy Matrix}
\label{subsec:outeralgo}

The algorithm as stated has an iteration complexity that's $O(k^{3/2}\epsilon^{-5/2})$,
which is small enough for our purposes.

However, a direct implementation of the generation of the energy matrix $\pd(e)$
requires looping through each of the $2^k$ sign vectors $\signv \in \signlist$
and computing $\signv^T \flowv$, which takes time exponential in $k$.

To alleviate this problem, note that the requirement of Theorem \ref{thm:roughweights}
allows us to compute the matrix for some set of weights $\approxweightv$ where
$\weight(\signv) \leq \approxweightv \leq (1 + \epsilon) \weight(\signv)$.
Specifically we show the following:

\begin{theorem}
\label{thm:approxenergy}
Given a flow $\flowv$ such that $||\flowv||_1 \leq \rho' \capacity$,
there is an algorithm \textsc{Energy} that computes a matrix $\pd$ where

\begin{align}
\pd
= & \sum_{\signv} \approxweightv \signv\signv^T \\
\exp \left(\frac{\signv^T \flowv}{\capacity} \right)
\leq & \approxweightv
\leq (1 + \epsilon) \exp \left(\frac{\signv^T \flowv}{\capacity} \right)
\end{align}

In $\tilde{O}(k^4 \rho' \epsilon^{-1})$ time.
\end{theorem}

Since we have the signs of each of the $\flow_i$,
we can easily find the value $\signv$ that maximizes $\signv^T \flowv$.
We let this set of signs be $\optsignv$, then we have for all $\signv \in \signlist$:

\begin{align}
\exp \left(\frac{\signv^T \flowv}{\capacity} \right)
&= \exp \left(\frac{\optsignv^T \flowv}{\capacity} \right)
\exp \left(- \frac{(\optsignv - \signv)^T \flowv}{\capacity} \right)
\end{align}

The first term is a constant, therefore it suffices to get good approximations
for the second term.
To do so we round each entry of $\flowv$ to the lowest integral multiple of
$\frac{\epsilon}{k}$ and bound the error as follows:

\begin{lemma}
Let $\approxflowv$ be $\flowv$ with each entry rounded towards $0$ to the nearest
multiple of $\frac{\epsilon}{3k}\capacity$.
Then we have:

\begin{align}
\exp \left(- \frac{(\optsignv - \signv)^T \flowv}{\capacity} \right)
\leq & \exp \left(- \frac{(\optsignv - \signv)^T \approxflowv}{\capacity} \right)   \nonumber \\
\leq & (1 + \epsilon) \exp \left(- \frac{(\optsignv - \signv)^T \flowv}{\capacity} \right)
\end{align}

Also, $||\approxflowv||_1 \leq \rho' \capacity$ as well.

\end{lemma}

\Proofof{Theorem \ref{thm:approxenergy}}
Note that by the choice of $\optsignv$, $\optsign_i \flow_i \geq 0$ in each of commodity
$i$.
Therefore $(\optsignv - \signv)_i \flowv_i$ is either $0$ or $2 \flowv_i$.
By the rounding rule we have:

\begin{align}
|\flowv_i| - \frac{\epsilon}{3k} \capacity
\leq & |\approxflowv_i|
\leq |\flowv_i|
\end{align}

Which gives us the bound on $||\approxflowv||_1$.
When combined with the fact that $\approxflowv_i$ having the same sign as $\flowv_i$ gives:

\begin{align}
(\optsignv - \signv)_i \flowv_i - \frac{2\epsilon}{3k} \capacity
\leq & (\optsignv - \signv)_i \approxflowv_i
\leq (\optsignv - \signv)_i \flowv_i
\end{align}

Summing this over the $k$ commodities gives:

\begin{align}
(\optsignv - \signv)^T \flowv - \frac{2\epsilon}{3} \capacity
\leq & (\optsignv - \signv)^T \approxflowv
\leq (\optsignv - \signv)^T \flowv
\end{align}

Exponentiating both sides of Fact \ref{fact:log} Part \ref{part:loglower} gives
$\exp(\frac{2 \epsilon}{3}) \leq (1 + \epsilon)$,
from which the result follows.
\QED

After this rounding, the values of $\signv^T \approxflowv$ can only be multiples of
$\frac{\epsilon}{k} \capacity$ between $[-\rho' \capacity, \rho' \capacity]$.
This allows us to narrow down the number of possible values of
$\exp(\frac{\signv^T\approxflowv}{\capacity})$ to one of $O(\rho' k\epsilon{-1})$
values.
Further more, notice that to calculate $\pd_{ij}$
it suffices to find the list of values of $\frac{\signv^T\approxflowv}{\capacity}$
among all $\signv$ such that $\signv_i = \signv_j$ and the list where $\signv_i \neq \signv_j$.
Each of these calculations can be done in $\tilde{O}(\rho' k^2\epsilon^{-1})$ time using
the following lemma:

\begin{lemma}
\label{lem:repeatedfft}
Given a list of positive integer values $a_1, a_2, \ldots a_k$
such that $\sum_{i} a_i = N$, there is an algorithm
\textsc{ConvolveAll} that computes in $\tilde{O}(Nk\log^2{N})$ time
for each $j \in [1, N]$ the number of subsets
$S \subseteq \{1, 2, \ldots k\}$ such that $\sum_{i \in S} a_i = j$.
\end{lemma}

\Proof
Since the ordering is irrelevant, we may assume that
$a_1 \leq a_2 \leq \ldots a_k$.
Then there exist an index $i$ such that $\sum_{1 \leq j \leq i} a_j$
and $\sum_{i+1 \leq j \leq k-1} a_j$ are both at most $N/2$.
Suppose we have two lists containing the number of sums for each
value between $0$ and $N/2$, then taking their convolution can be
done in $O(N\log{N})$ multiplications involving $k$ bit numbers
(\cite{CLRS01} chapter 30).
The last entry of $a_k$ can be incorporated similarly.
This leads us to the following recurrence on $T(N)$,
the time required to compute the answer when the total sum is $N$:
\begin{align}
T(N) \leq 2T(N/2) + \tilde{O}(Nk\log{N})
\end{align}
Solving gives $T(N) = \tilde{O}(Nk \log^2{N})$.
\QED

\begin{algo}[h]\label{alg:approxenergy}
\qquad

\textsc{Energy}
\vspace{0.05cm}

\underline{Input:}
A $k$ commodity flow $\flowv$.
Capacity $\capacity$, parameter $\rho'$ such that $||\flowv||_1 \leq \rho' \capacity$.
Error bound $\epsilon$.

\underline{Output:} Approximate energy matrix satisfying the guarantees
of Theorem \ref{thm:approxenergy}

\vspace{0.2cm}

\begin{algorithmic}[1]
\STATE{Compute the set of signs that maximizes $\signv^T \flowv$, $\optsignv$}
\FOR{$i = 1 \ldots k$}
    \IF{$\flowv_i \geq 0$}
        \STATE{$\approxflowv_i = \capacity \frac{\epsilon}{k} \lfloor \frac{\flowv_i} k \epsilon^{-1}\rfloor$}
    \ELSE
        \STATE{$\approxflowv_i = -  \capacity \frac{\epsilon}{k} \lfloor - \frac{\flowv_i} k \epsilon^{-1}\rfloor$}
    \ENDIF
\ENDFOR
\FOR{$i = 1 \ldots k$}
    \FOR{$j = 1 \ldots k$}
        \FOR{Each setting of $\signv_i$, $\signv_j$}
            \STATE{For each $l \neq i, j$, create $a_l = 2\frac{\approxflow_l}{\capacity} k \epsilon^{-1}$}
            \STATE{$\vecb \leftarrow \textsc{ConvolveAll}(\veca)$}
            \FOR{$-\rho' k\epsilon^{-1} \leq l \leq \rho' k \epsilon^{-1}$}
                \STATE{$\pd_{ij} \leftarrow \pd_{ij} +
                    \vecb_l \cdot \signv_i \cdot \signv_j \cdot
                        \exp(\frac{\epsilon}{\rho} \frac{\optsignv^T\flowv}{\capacity} - l)$}
            \ENDFOR
        \ENDFOR
	\ENDFOR
\ENDFOR
\RETURN {$\pd$}
\end{algorithmic}

\caption{Algorithm for computing approximate energy matrix}\label{alg:approxenergy}
\end{algo}

The overall pseudocode for computing this energy matrix is shown in Algorithm \ref{alg:approxenergy}.
Summing over all $O(k^2)$ entries gives the total running time.
It's worth noting that because the matrix entries consists of differences of weights,
the matrix that we obtain can have some entries that are very different than what we
would obtain if we use the exact values of $\weight(\signv)$.
Their similarity is obtained through the similarity of $\tilde{w}$ and $\hat{w}$
as they are weights on positive semi-definite outer products.

We can now bound the overall running time of the algorithm:

\begin{corollary}
\textsc{MaxConcurrentFlow1} runs in $\tilde{O}(k^{3/2}\epsilon^{-5/2})$ iterations,
where each iteration takes time
$\tilde{O}(m^{4/3}k^{5/2}\epsilon^{-19/6} + m k^{5}\epsilon^{-3/2})$,
for a total running time of $\tilde{O}(m^{4/3}k^{4}\epsilon^{-17/3} + m k^{13/2} \epsilon^{-4})$.
\end{corollary}

\Proof
The iteration count from \ref{thm:roughweights} completes the proof.
The first term follows from Theorem \ref{thm:multiplicativeweights}
and $\kappa(\pd(e)) \leq \sqrt{k} \epsilon^{-1/2}$,
Theorem \ref{thm:roughweights} Part \ref{part:convergence} gives that
$||\flowv^{(t)}||_1 \leq O(\numiter)$, which gives
$||\frac{\epsilon}{\rho} \flowv^{(t)}||_1 \leq \frac{\epsilon \numiter}{\rho} = \tilde{O}(k)$.
Letting $\rho' = \tilde{O}(k)$ in Theorem \ref{thm:approxenergy} then gives the second term
in the bound.
\QED

\section{Comments/Extensions}
\label{sec:comments}

We have shown an approach of dealing with the coupling of the $k$
commodities by associating an energy matrix with them.
This allows us to approximate multicommodity flows in time
$\tilde{O}(m^{4/3}\poly(k, \epsilon^{-1}))$.
We believe that our approach is quite general and extends
naturally to other couplings between sets of $k$ flows/vertex
labels, with the most natural generalization being Markov
random fields \cite{KS80, SZSVKATR08}.
Since reductions from multicommodity flow to $\tilde{O}(k\epsilon^{-2})$
calls of minimum cost flows are known \cite{LeightonMPSTT91}.
A stronger result would be an approximation of minimum cost flow that runs
in $\tilde{O}(m^{4/3}\poly(k, \epsilon^{-1})$ time.
This problem is significantly harder due to it incorporating both $L_\infty$
and $L_1$ constraints.
Therefore, it's likely that a more intricate set of energy matrices
is needed to proceed in this direction.

When viewed from the perspective of combinatorial preconditioning,
we were able to solve multicommodity electrical flows by
adding $\epsilon$ slack to the edges, thus 'fixing' the condition number.
For the purpose of obtaining $1 \pm \epsilon$ approximations to
combinatorial problems, this does not modify the solution by too much.
However, it is unlikely to be applicable inside algorithms
whose dependency on $\epsilon$ is $O(\log(1/\epsilon))$,
as very little slack can be added onto the edges.
As a result, we believe that obtaining fast solvers for the class of matrices
that arise from quadratically coupled flows is an interesting direction
for future work.




\begin{spacing}{0.7}
  \begin{small}
    \bibliographystyle{alpha}
    \newcommand{\etalchar}[1]{$^{#1}$}

  \end{small}
\end{spacing}

\begin{appendix}

\section{Obtaining Almost Optimal Electrical Flow from Potentials}
\label{sec:solvererror}

This approximate solve then allows us to compute electrical flows in a way
analogous to theorem 2.3 of \cite{ChristianoKMST10}.
Pseudocode of the algorithm is shown in Algorithm \ref{alg:oracle}.

\begin{algo}
\qquad

\textsc{QuadraticallyCoupledFlow}
\vspace{0.05cm}

\underline{Input:}
Graph $G=(V,E)$
with corresponding edge-vertex incidence matrix $\edgevertex$.
Demand vector $\demandv$
Energy matrices for the edges $\pd$ along with bounds $U$, $\kappa$ such that
    $I \preceq \pd \preceq U \cdot I$ and
    $\lambda_{\max}(\pd(e)) \preceq \kappa \lambda_{\min}(\pd(e))$.
Error bound $\delta$.

\underline{Output:}
Approximate vertex potentials $\approxpotentialvertv$ and
flow $\approxflowv$ that satisfies the constraints of Theorem

\vspace{0.2cm}

\begin{algorithmic}[1]
    \STATE{Let $\laplacianglobal$ be an implicit representation of the matrix $\edgevertexglobal \pd \edgevertexglobal$}
    \STATE{Use \textsc{PreconCheby} to find $\approxpotentialunscaledv = \laplacianglobal^{+} \demandv$
        with error of $\epsilon = \frac{\delta}{5m^6k^2\maxratio^4}$}
    \STATE{$\approxscale \leftarrow \sqrt{\approxpotentialunscaledv ^T \laplacianglobal \approxpotentialunscaledv}$}
    \STATE{$\approxpotentialvertv = \frac{1}{\lambda} \approxpotentialunscaledv$}
    \STATE{Compute ohmic flow $\ohmicflowv = \pd^{-1} \edgevertexglobal \approxpotentialunscaledv$}
    \FOR{$i = 1 \ldots k$}
        \STATE{$\approxflowv_i \leftarrow \textsc{MakeKirchhoff}(G, \demandv_i, \ohmicflowv_i)$}
    \ENDFOR
    \RETURN{$\approxpotentialvertv, \approxflowv$}
    \caption{Algorithm for computing near-optimal electrical potential/flow pair}
    \label{alg:oracle}
\end{algorithmic}

\end{algo}

The algorithm makes calls to $\textsc{MakeKirchoff}$, which converts a flow
$\ohmicflowv$ that doesn't satisfy conservation of flow at vertices in $W$
to one that does without too much increase in congestion.
Its properties are proven in Lemma \ref{lem:makekirchhoff}.

\begin{algo}
\qquad

\textsc{MakeKirchhoff}
\vspace{0.05cm}

\underline{Input:}
Graph $G=(W \subseteq V,E)$ with demands $\demandv$.
A Ohmic flow $\flowv$.

\underline{Output:}
A flow that satisfies conservation of flow at all vertices in $W$

\vspace{0.2cm}

\begin{algorithmic}[1]
\STATE{$\approxflowv \leftarrow \flowv$}
\STATE{Pick any vertex $r$}
\STATE{compute a spanning tree, $T$}
\FOR{each vertex $u \neq r$}
    \STATE{route $\demand_u - (\edgevertexglobal \approxflowv)_u$ units of flow from $u$ to $r$ along $T$}
\ENDFOR
\RETURN{$\approxflowv$}
\caption{Algorithm for making an arbitrary flow Kirchhoff}\label{alg:krichoff}
\end{algorithmic}

\end{algo}

We start by bounding the errors incurred by \textsc{MakeKirchhoff}

\begin{lemma}
\label{lem:makekirchhoff}
Given a capacitated graph $G = (V, E, \capacityv)$, a set of demands $\demandv$
a flow $\ohmicflowv$ such that $||\edgevertexglobal^T \ohmicflowv||_\infty \leq \gamma$.
\\$\textsc{MakeKirchhoff}(G, \demandv, \ohmicflowv)$ returns a flow $\tilde{f}$ such that:
\begin{enumerate}
    \item \label{part:demands} $\edgevertexglobal^T  \approxflowv = \demandv$.
    \item \label{part:extracongestion} For all edges $e \in E$, $||\ohmicflow_e - \approxflow_e||_\infty \leq m \gamma$.
\end{enumerate}
\end{lemma}

\Proof
Part \ref{part:demands} follows from the construction for each vertex $u \neq r$
and $\onesv^T \demandv = 0$.
To bound the extra congestion, note that each edge in the tree is used in at most $n < m$
paths that route at most $\gamma$ units of flow each.
This gives an extra congestion of at most $m \gamma$ and therefore Part \ref{part:extracongestion}.
\QED

\Proofof{Theorem \ref{thm:approxelectrical}}

We start off by showing that $\approxscale$ does not differ by much from
$\optscale = \sqrt{\optpotentialvertv^T \laplacianglobal \optpotentialvertv}$:

\begin{lemma}
\label{lem:objdifference}
\begin{align*}
|\approxscale - \optscale|
\leq & \epsilon \optscale
\end{align*}
\end{lemma}

\Proof
\begin{align}
|\approxscale - \optscale|
= & | ||\approxpotentialunscaledv||_\laplacianglobal - ||\optpotentialunscaledv||_\laplacianglobal | \nonumber \\
\leq & ||\approxpotentialunscaledv - \optpotentialunscaledv||_\laplacianglobal
    \qquad \mbox{Since $||\cdot||_\laplacianglobal$ is a norm} \nonumber \\
\leq & \epsilon ||\optpotentialunscaledv||_\laplacianglobal
= \epsilon \optscale
\end{align}
\QEDpart{Lemma \ref{lem:objdifference}}

Algebraic manipulations of this give:

\begin{corollary}
    \begin{align}
        (1 - \epsilon) \optscale \leq & \approxscale \leq (1+\epsilon) \optscale \\
        (1 - 2\epsilon) \frac{1}{\optscale} \leq & \frac{1}{\approxscale} \leq (1+2\epsilon) \frac{1}{\optscale}
    \end{align}
\end{corollary}

This allows us to show that $\approxflowv = \approxscale \pd^{-1} \edgevertexglobal \approxpotentialunscaledv$
does not differ from $\optflowv = \optscale \pd^{-1} \edgevertexglobal \optpotentialunscaledv$ by too much:

\begin{lemma}
\label{lem:totaldifference}
\begin{align}
||\ohmicflowv - \optflowv||_{\pd}
\leq & \sqrt{(\approxflowv - \optflowv)^T \pd (\approxflowv - \optflowv)} \nonumber \\
\leq & 4 \epsilon \optscale^2 \nonumber \\
= & 4 \epsilon \eflow_{\optflowv}(\pd)
\end{align}
\end{lemma}

\Proof
\begin{align}
||\ohmicflowv - \optflowv||_{\pd}
= & ||\pd^{-1} \edgevertexglobal \approxpotentialunscaledv
    - \pd^{-1} \edgevertexglobal \optpotentialunscaledv||_{\pd} \nonumber \\
= & ||\approxpotentialunscaledv - \optpotentialunscaledv||_{\laplacianglobal}
    \qquad \mbox{By definition of $\laplacianglobal$} \nonumber \\
\leq & \epsilon ||\optpotentialunscaledv||_{\laplacianglobal}
\end{align}
\QEDpart{Lemma \ref{lem:totaldifference}}

This in turn gives bounds on the maximum entry-difference
of $\ohmicflowv$ and $\optflowv$:

\begin{corollary}
\label{cor:maxdifference}
\begin{align*}
||\approxflowv - \optflowv||_\infty \leq \epsilon \optscale
\end{align*}
\end{corollary}

\Proof
Follows from $\pd \succeq \identity$
and $||\vecv||_\infty \leq ||\vecv||_2$ for any vector $\vecv$.
\QEDpart{Corollary \ref{cor:maxdifference}}

Since $\optflowv$ meets the demands at each vertex, the amount
that $\ohmicflowv$ does not meet the demand by is at most
$\epsilon mk \optscale$.
Lemma~\ref{lem:makekirchhoff} gives that the difference along each
edge after running \textsc{MakeKirchhoff} is at most $\epsilon m^2k^2 \optscale$.
The following two bounds are immediate consequences of this:

\begin{enumerate}
\item
\begin{align}
||\approxflowv - \optflowv||_{\pd}
\leq & ||\approxflowv - \ohmicflowv||_{\pd} + ||\ohmicflowv - \optflowv||_{\pd} \nonumber \\
\leq & \epsilon \maxratio m^2k^2 \optscale + \epsilon \optscale
    \qquad \mbox{Since $\pd \preceq \maxratio \identity$} \nonumber \\
\leq & 2 \epsilon \maxratio^2 m^2 k^2 \optscale
\end{align}

\item
\begin{align}
\left| \sqrt{\eflow_{\approxflowv}(\pd)} - \sqrt{\eflow_{\optflowv}(\pd)}\right|
= & \left| ||\approxflowv||_\pd - ||\optflowv||_{\pd} \right|  \nonumber \\
\leq & ||\approxflowv - \optflowv||_\pd
\leq 2 \epsilon \maxratio^2 m^2 k^2 \optscale
\end{align}

Which in turn implies that
$
\sqrt{\eflow_{\approxflowv}(\pd)}
\leq (1 + 5 \epsilon \maxratio^2 m^2 k^2) \sqrt{\eflow_{\optflowv}(\pd)}
$
when $\epsilon \maxratio^2 m^2 k^2 < 0.01$.

\end{enumerate}

This enables us to derive the overall bound for the $L_2$ energy
difference of the two flows.
We first need one more identity about the pointwise difference
$| \eflow_{\approxflow}(\pd, e) - \eflow_{\optflow}(\pd, e)|$.
Since ${\pd(e)}$ is positive definite, we may write it as
$\matq(e)^T\matq(e)$.
Then the difference in energy can be written as:

\begin{align}
& \left | \eflow_{\approxflow}(\pd, e) - \eflow_{\optflow}(\pd, e) \right| \nonumber \\
= & \left| ||\matq(e) \approxflow(e)||_2^2 - ||\matq(e) \optflow(e)||_2^2 \right| \nonumber \\
= & \left| (\matq(e) \approxflow(e) + \matq(e) \optflow(e))^T
(\matq(e) \approxflow(e) - \matq(e) \optflow(e)) \right|
\qquad \text{By applying $a^2-b^2=(a+b)(a-b)$ entry-wise} \nonumber \\
\leq & ||\matq(e) \approxflow(e) + \matq(e)  \optflow(e)||_2 
		||\matq(e) \approxflow(e) - \matq(e)  \optflow(e)||_2 
\qquad \mbox{by the Cauchy-Schwarz inequality} \nonumber \\
= & ||\approxflow(e) + \optflow(e)||_{\pd(e)}
	||\approxflow(e) - \optflow(e)||_{\pd(e)}
\end{align}

Since $\pd$ is a block diagonal matrix divided by the edges, we have
$||\vecv||_\pd = \sum_e ||\vecv(e)||_{\pd(e)}$ for any vector $\vecv$.
Therefore:
\begin{align}
||\approxflow(e) + \optflow(e)||_{\pd(e)}
\leq & ||\approxflow + \optflow||_\pd \nonumber \\
\leq & \sqrt{4 \eflow_{\approxflowv}(\pd) }
	\qquad \mbox{since $(a+b)^2 \leq 2(a^2+b^2)$
    and $\eflow_{\optflowv}(\pd) \leq \eflow_{\approxflowv}(\pd)$} \\
||\approxflow(e) - \optflow(e)||_{\pd(e)}
\leq & ||\approxflow - \optflow||_\pd \nonumber \\
\leq & (1 + 5 \epsilon \maxratio^2 m^2 k^2) \optscale 
\end{align}

Combining them gives:

\begin{align}
\left| \eflow_{\approxflow}(\pd, e) - \eflow_{\optflow}(\pd, e) \right|
\leq &  \left( 5 \epsilon \maxratio^2 m^2 k^2 \optscale \right)
\left( (1 + 5 \epsilon \maxratio^2 m^2 k^2) \optscale  \right) \nonumber \\
\leq & 10 \epsilon \maxratio^2 m^2 k^2 \eflow_{\optflowv}(\pd)
	\qquad \mbox{When $5 \epsilon \maxratio^2 m^2 k^2 < 1$}
\end{align}

It can be checked that when $\delta < 0.1$, setting
$\epsilon = \frac{\delta}{2 m^2 k^2 \maxratio}$ satisfies the condition
for the last inequality, as well as bounding the last term by $\delta$.
For this setting, we have $\log(1/\epsilon) \leq O(\log(\maxratio m k / \delta))$.

\QEDpart{Theorem \ref{thm:approxelectrical}}

\section{Maximum Weighted Multicommodity Flow}
\label{sec:weighted}
The maximum weighted multicommodity flow problem is a related problem that
maximizes $\sum_i \lambda_i \flowvalue_i$ for a series of weights
$\lambda \in \Re^{k}_+$, where $\flowvalue_i$ is the amount of flow
of commodity $i$ that's routed between the demand pairs.
In this section we give an overview of how to extend our algorithm to this problem.
However, in order to simplify presentation we assume that the solves are exact.
Errors from these solves can be analyzed in steps similar to those in our
algorithm for maximum concurrent flow.

In our notation, the problem can be formulated as
finding flows $\flowv_1 \ldots \flowv_k$ such that
$\edgevertex \flowv_i = \flowvalue_i \demandv_i$ and
maximizing $\sum_i \lambda_i \flowvalue_i$.
Furthermore it can be reduced to the decision problem of finding
$\flowv_i$, $\flowvalue_i$ such that $\sum_i \flowvalue_i = 1$.
Note that due to the undirected nature of the flow, we can allow $\flowvalue_i$
to be negative since $\lambda_i \geq 0$ means negating the flow
of the $i^{th}$ commodity
can only improve the overall objective.
Once we introduce the $k$ additional variables $\flowvalue_1 \ldots \flowvalue_k$,
the maximum weighted multicommodity flow problem can be formulated as:

\begin{align*}
\text{maximize: } & \sum_i \flowvalue_{i} & &\\
\text{subject to: } &\sum_{i = 1}^k |\flowv_i(e)| \leq u(e) & \forall e \in E\\
& \edgevertex ^T \flowv_i = \flowvalue_i \demandv_i & \forall 1 \leq i \leq k
\end{align*}

By binary search and appropriate scaling of $\demandv_i$, it suffices to
check whether there is a solution $(\flowv, \flowvaluev)$ where
$\onesv^T \flowvaluev = 1$.
Furthermore we can use $\demandmat$ to denote the $kn \times k$ matrix
with column $i$ being $\demandv_i$ extended to all $kn$ vertex/commodity
pairs.
Then if we let $\flowvaluev$ denote the vector containing all the flow values,
the corresponding quadratically coupled flow problem becomes:

\begin{align*}
\text{minimize: } & \eflow(\pd, \flowv) & &\\
\text{subject to: } & \edgevertexglobal ^T \flowv = \demandmat \flowvalue\\
& \onesv^T \flowvaluev = 1
\end{align*}

As before, we let $\laplacianglobal = \edgevertexglobal^T \pd \edgevertexglobal$
and define the following quantities:

\begin{align}
\lambda
= & \frac{1}{\onesv^T (\demandmat \laplacianglobal^+ \demandmat)^+ \onesv}
\label{eq:deflambda} \\
\optflowvaluev
=& \lambda (\demandmat \laplacianglobal^+ \demandmat)^+ \onesv
\label{eq:defoptflowvaluev} \\
\optpotentialvertv
=& \laplacianglobal^+ \demandmat \optflowvaluev
\label{eq:defoptpotentialvertv} \\
\optflowv
=& \pd^{-1} \edgevertexglobal \optpotentialvertv
\label{eq:defoptflowv}
\end{align}

Note that the matrix $\demandmat \laplacianglobal^+ \demandmat$ can be
explicitly computed using $k^2$ solves, and the resulting $k$-by-$k$ matrix
can be inverted using direct methods in $O(k^{\omega})$ time.
The following lemmas similar to the ones about quadratically coupled flows
shown in Section \ref{sec:electrical} can be checked in an analogous way.

\begin{lemma}
\label{lem:weightedflowcon}
\begin{align*}
\edgevertexglobal^T \optflowv = \demandmat \optflowvaluev
\end{align*}
\end{lemma}

\Proof
\begin{align}
\edgevertexglobal^T \optflowv
= & \edgevertexglobal^T \pd^{-1} \edgevertexglobal \optpotentialvertv \nonumber \\
= & \laplacianglobal \optpotentialvertv \nonumber \\
= & \laplacianglobal \laplacianglobal^+ \demandmat \optflowvaluev \nonumber \\
= & \demandmat \optflowvaluev
\end{align}

\QED

\begin{lemma}
\label{lem:weightedoptvalue}
\begin{align*}
\eflow(\pd, \optflowv) = \epotential(\optpotentialvertv) = \lambda
\end{align*}
\end{lemma}

\Proof

\begin{align}
\eflow(\pd, \optflowv) 
= & \optflowv^T \pd \optflowv \nonumber \\
= & \optpotentialvertv^T \edgevertexglobal^T \pd^{-1}
	\pd \pd^{-1} \edgevertexglobal \optpotentialvertv \nonumber \\
= & \optpotentialvertv^T \laplacianglobal \optpotentialvertv \nonumber \\
= &  \optflowvaluev \demandmat^T \laplacianglobal^+
	\laplacianglobal \laplacianglobal^+ \demandmat \optflowvaluev \nonumber \\
= &  \optflowvaluev \demandmat^T \laplacianglobal \demandmat \optflowvaluev \nonumber\\
= & \lambda^2
	 \onesv^T (\demandmat \laplacianglobal^+ \demandmat)^+ 
	 (\demandmat^T \laplacianglobal \demandmat)
	 (\demandmat \laplacianglobal^+ \demandmat)^+ \onesv \nonumber \\
= & \lambda^2 \frac{1}{\lambda} = \lambda
\end{align}

\QED

\begin{lemma}
\label{lem:weightedopt}
For any $(\flowv, \flowvaluev)$ pair that satisfies
$\edgevertexglobal^T \flowv = \demandmat \flowvaluev$
and $\onesv^T \flowvaluev = 1$, we have:
\begin{align*}
\eflow(\pd, \flowv) \geq \lambda
\end{align*}
\end{lemma}

\Proof
Since $ \epotential(\optpotentialvertv) = \lambda$ by Lemma \ref{lem:weightedoptvalue},
it suffices to show
$\eflow(\pd, \flowv) \epotential(\optpotentialvertv) \geq \lambda^2$.

\begin{align}
\eflow(\pd, \flowv) \epotential(\optpotentialvertv)
= & (\flowv^T \pd \flowv )
((\edgevertexglobal \optpotentialvertv)^T \pd^{-1}
(\edgevertexglobal \optpotentialvertv))
	\qquad \text{By definition of $\flowv$ in Equation \ref{eq:defoptflowv}}
\nonumber \\
\geq & (\flowv^T \edgevertexglobal \optpotentialvertv)^2
	\qquad \text{By Cauchy-Schwarz inequality} \nonumber \\
= & (\flowvaluev^T \demandmat^T \optpotentialvertv)^2
	\qquad \text{Since $\edgevertexglobal^T \flowv = \demandmat \flowvaluev$}
\nonumber \\
= & (\flowvaluev^T \demandmat^T \laplacianglobal^+ \demandmat \optflowvaluev)^2
	\qquad \text{By definition of $\optpotentialvertv$ in Equation \ref{eq:defoptpotentialvertv}}
\nonumber\\
= &  (\flowvaluev^T (\lambda\onesv))^2
	\qquad \text{By definition of $\optflowvaluev$ in Equation \ref{eq:defoptflowvaluev}}
\nonumber \\
= & \lambda^2
	\qquad \text{Since $\onesv^T \flowvaluev = 1$}
\end{align}
\QED

Furthermore, Lemma \ref{lem:potentials}, Part \ref{part:potentialincrease},
which is crucial for showing an increase in the
minimum energy of the quadratically coupled flow, still holds.
Specifically, Equations \ref{eq:erfraction} and \ref{eq:newer} adapts readily
with the $\potentialvertv$.
Therefore generalizing Theorem \ref{thm:multiplicativeweights} and combining
it with Theorem \ref{thm:outermatrixexp}
gives an analogous result for approximating the maximum weighted
multicommodity flow problem:

\begin{theorem}
\label{thm:maxweighted}
Given an instance of the maximum weighted multicommodity flow problem
with $k$ commodities on a graph with $m$ edges and an error parameter
$\epsilon > 0$.
A solution with weight at least $(1 - \epsilon)$ of the maximum can be produced
in $\tilde{O}(m^{4/3}\poly(k, \epsilon^{-1}))$ time.
\end{theorem}

\end{appendix}

\end{document}